\documentstyle[prb,epsfig,eqsecnum,aps,twocolumn]{revtex}

\newfont{\smsbm}{msbm10 at 9pt}
\newfont{\msbm}{msbm10 at 12pt}

\begin{document}
\draft

\makeatletter \global\@specialpagefalse

\twocolumn[{\hsize\textwidth\columnwidth\hsize\csname
@twocolumnfalse\endcsname
\begin{tabbing}
\=222222222222222222\= 222222222 \=222222\= \kill
 \>{\epsfxsize=2cm \epsffile{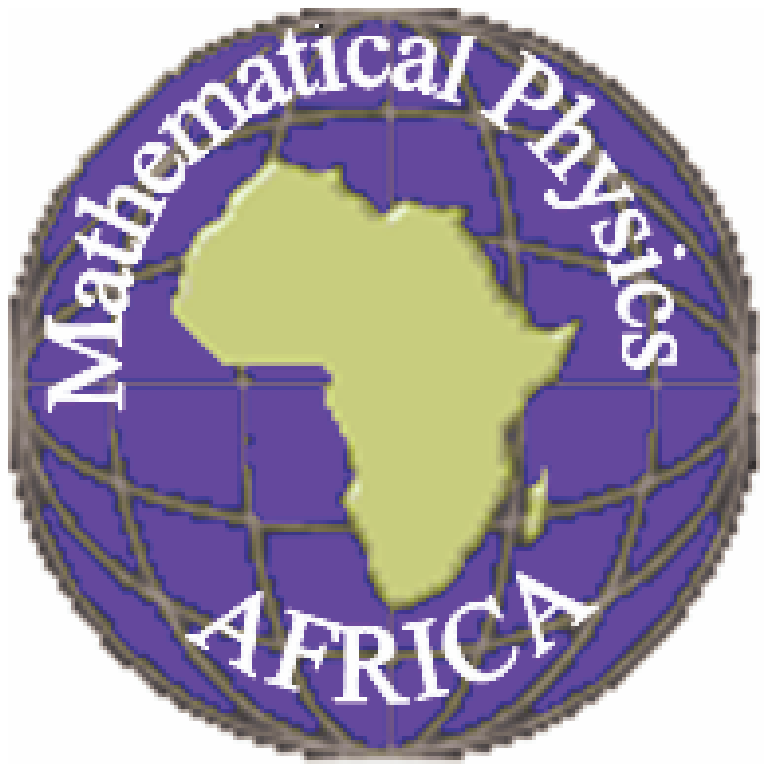}}\> \>{}\>{}
\\ \>{} \>{}\>{}\>{}
\\ \>
\end{tabbing}

 \title{{\vspace{-5cm}\rm\small
 \rightline {\underline{\it{ African Journal Of Mathematical Physics {\bf{1}} (2004)1-19}}}\vspace{1cm}}
  \vskip1truecm \Large\bf Th\'{e}orie-M sur des Vari\'{e}tes d'holonomie G2}
 \vskip 2 cm
\author{ Lalla Btissam Drissi $^{1,2}${\vspace{0.5cm}}\\
{\small  \it 1.Lab/UFR Physique des Hautes Energies, Facult\'{e}
des Sciences, Rabat, Morocco.\vspace{0.1cm}}
\\ {\small \it 2. Groupement National de Physique des Hautes
Energies, GNPHE; Si\`{e}ge focal, Rabat, Morocco.\vspace{0.1cm}}}
 \maketitle

\bigskip \bigskip
\centerline{\bf Abstract}
 \begin{abstract}
L'un des objectifs de ce papier consiste \`{a} ramener la
th\'{e}orie-M, consid\'{e}r\'{e}e comme la limite des cinq
diff\'{e}rentes th\'{e}ories des supercordes, \`{a} une
th\'{e}orie \`{a} $3+1$ dimensions et avec un nombre de
supercharges minimal. Nous montrons que ce mod\`{e}le ne se
r\'{e}alise qu'\`{a} travers l'introduction d'une vari\'{e}t\'{e}
tr\`{e}s sp\'{e}ciale d'holonomie $\mathbf{G}_{2}$. Par la suite,
nous proc\'{e}dons \`{a} l'\'{e}tude des mod\`{e}les
r\'{e}sultants de la compactification de la th\'{e}orie-M sur
cette vari\'{e}t\'{e} dans ses deux cas: r\'{e}gulier et
singulier. \vspace{0.25cm}
 \end{abstract}}
  \bigskip \bigskip] \setcounter{page}{1}

\section{Introduction}

\makeatletter \global\@specialpagefalse

\thispagestyle{empty} L'ultime but de la physique des hautes \'{e}nergies
est de fournir une th\'{e}orie unique qui d\'{e}crive l'univers dans son
ensemble \footnote{{\scriptsize {{AJMP} is an African Scientific Journal
published temporary by Moroccan Grouping In High Energy Physics, Rabat, {\it %
e-mail: ajmp@fsr.ac.ma }}}}. Cette qu\^{e}te fut lanc\'{e}e dans les ann\'{e}%
es 1960 sous l'impulsion des physiciens Glashow, Salam et Weinberg. Ces
trois chercheurs sont parvenus \`{a} unifier les deux interactions faible et 
\'{e}lectromagn\'{e}tique en faisant appel \`{a} des sym\'{e}tries internes.
D\`{e}s lors, cette th\'{e}orie fut connue sous le nom de ''th\'{e}orie \'{e}%
lectrofaible''. Parall\`{e}lement \`{a} cette derni\`{e}re, le ''Mod\`{e}le
Standard'' est apparu pour tenter d'expliquer et d'ordonner la grande quantit%
\'{e} de particules d\'{e}couvertes. Toutefois, vers les ann\'{e}es soixante
dix, cette branche de la physique a pu d\'{e}crire, dans ce qu'on appelle la
th\'{e}orie de la grande unification, les trois forces \'{e}lectromagn\'{e}%
tique, faible et forte. Mais malgr\'{e} son succ\`{e}s remarquable, le Mod%
\`{e}le Standard demeure insatisfaisant puisqu'il n'inclut pas la gravit\'{e}%
. Ainsi dans le but de faire face \`{a} cet obstacle, d'autres mod\`{e}les
ont \'{e}t\'{e} d\'{e}velopp\'{e}s parmi eux la th\'{e}orie des cordes.%
\newline
Le postulat de base de la th\'{e}orie des cordes consiste \`{a} repr\'{e}%
senter la particule non pas par un point, mais plut\^{o}t par une corde dot%
\'{e}e d'une longueur tr\`{e}s petite \cite{ref0,ref1,ref2}. Celle-ci balaye
un chemin qui est une surface bidimensionnelle conforme.\newline
La th\'{e}orie des cordes la plus simple, est celle des cordes bosoniques
qui semble n'\^{e}tre valable que si l'espace-temps poss\`{e}de 26
dimensions au lieu de nos quatre habituelles! En plus de sa dimension
critique, cette th\'{e}orie n'est pas consistante en raison de quelques
lacunes qu'elles pr\'{e}sentent (voir section II). Afin d'y rem\'{e}dier, la
th\'{e}orie des supercordes ou des cordes supersym\'{e}triques fut
introduite. Cette derni\`{e}re est cens\'{e}e constituer un jour le meilleur
espoir de d\'{e}velopper une ''th\'{e}orie du tout'' fondamentale, surtout
qu'elle contient dans son spectre une particule identifi\'{e}e au graviton (m%
\'{e}diateur des interactions gravitationnelles).\newline
Selon cette th\'{e}orie, il existe cinq types de supercordes \`{a} dix
dimensions deux de supersym\'{e}trie ${\cal N}=2$ et trois autres ayant une
supersym\'{e}trie ${\cal N}=1$ \cite{ref3}. Quoiqu'elles vivent dans un
espace-temps dont la dimension est inf\'{e}rieure \`{a} 26, celle-ci reste tr%
\`{e}s sup\'{e}rieure \`{a} nos quatre dimensions r\'{e}elles. Heureusement,
le proc\'{e}d\'{e} de compactification qui prend sa source des anciens
travaux de Kaluza-Klein, permet aux cinq mod\`{e}les d'\^{e}tre d\'{e}finis
dans des dimensions inf\'{e}rieures \cite{ref3,ref5}. Ce m\'{e}canisme offre
d'une part la possibilit\'{e} de r\'{e}duire la dimension, et d'autre part
celle de connecter les diff\'{e}rents mod\`{e}les obtenus par le biais des
sym\'{e}tries de dualit\'{e}s. Parmi ces sym\'{e}tries, nous citons la dualit%
\'{e}-T et aussi la dualit\'{e}-S qui permet de d\'{e}terminer la limite de
couplage fort de trois mod\`{e}les seulement des supercordes: type IIB, type
I et H\'{e}t\'{e}rotique $SO(32).$ Tandis qu'aux deux autres mod\`{e}les:
type IIA et H\'{e}t\'{e}rotique $E_{8}\times E_{8},$ il a \'{e}t\'{e}
conjectur\'{e} que ce sont des limites non perturbatives de la th\'{e}%
orie-M. Cette th\`{e}orie est d\'{e}crite \`{a} faible \'{e}nergie par une th%
\'{e}orie de supergravit\'{e} \`{a} onze dimensions ayant 32 supercharges
conserv\'{e}es (${\cal N}=1$).

La force de la th\'{e}orie-M c'est qu'entre autre elle est envisag\'{e}e
comme la limite des cinq diff\'{e}rentes th\'{e}ories des supercordes.
Cependant elle se propage, comme nous venons de le mentionner, dans un
espace \`{a} onze dimensions alors que l'espace r\'{e}el est seulement \`{a} 
$3+1$ directions non compactes. A cet \'{e}gard, l'un des objectifs centraux
de ce papier consiste \`{a} ramener la th\'{e}orie-M \`{a} une th\'{e}orie 
\`{a} $3+1$ dimensions et avec un nombre de supercharges minimales. Cette
derni\`{e}re ne s'obtient qu'\`{a} travers l'introduction d'une vari\'{e}t%
\'{e} tr\`{e}s sp\'{e}ciale qui doit v\'{e}rifier deux propri\'{e}t\'{e}s 
\`{a} la fois \cite{ref7,refA}:

\begin{enumerate}
\item Etre de dimension sept.

\item Avoir pour groupe d'holonomie le groupe ${\bf G}_{2}$ qui, lui seul,
permet de conserver le $\frac{1}{8}$ des supercharges initiales.
\end{enumerate}

Par cons\'{e}quent, seule la compactification de la th\'{e}orie-M sur cette
vari\'{e}t\'{e} assure l'obtention d'un mod\`{e}le \`{a} $D=4$ ${\cal N}=1 $
ayant un groupe de jauge ab\'{e}lien.

Compte tenu de la nouveaut\'{e} et de la grande importance de ce r\'{e}%
sultat trait\'{e}e dans une bonne partie de ce document, nous avons jug\'{e}
utile d'aller au del\`{a} de l'id\'{e}e de la compactification de la th\'{e}%
orie-M sur une vari\'{e}t\'{e} d'holonomie ${\bf G}_{2}$ r\'{e}guli\`{e}re.
Pour cette raison, nous nous sommes bas\'{e}s dans le d\'{e}veloppement de
cet aspect sur les travaux r\'{e}cents d'Acharya, Atiyah, Witten et d'autres 
\cite{ref7,ref15,ref16,ref17}. Selon ces recherches, l'attribution de r\'{e}%
alisations bien particuli\`{e}res \`{a} la vari\'{e}t\'{e} d'holonomie ${\bf %
G}_{2},$ localement au voisinage des singularit\'{e}s, est capable de nous
susciter un mod\`{e}le \`{a} quatre dimensions avec un nombre minimal de
supercharges et qui poss\`{e}de en plus un groupe de jauge non ab\'{e}lien
ainsi que de la mati\`{e}re chirale.

Pour d\'{e}velopper ces diff\'{e}rents axes nous proposons de r\'{e}partir
ce papier en trois sections que nous pr\'{e}sentons comme suit:

Dans la deuxi\'{e}me section, nous commen\c{c}ons par introduire la th\'{e}%
orie des cordes bosoniques tout en rappelant des notions de base de la th%
\'{e}orie des supercordes. Ensuite nous \'{e}voquons le proc\'{e}d\'{e} de
compactification ainsi que quelques exemples de sym\'{e}trie de dualit\'{e}.
Nous compl\'{e}tons cette section en traitant, selon deux approches diff\'{e}%
rentes, la th\'{e}orie-M comme l'une des cons\'{e}quences majeures de l'\'{e}%
tude de dualit\'{e}.

Quand \`{a} la troisi\`{e}me section, elle est consacr\'{e}e \`{a} la
compactification de la th\'{e}orie-M que nous \'{e}tudions en deux parties:

Dans la premi\`{e}re partie, nous pr\'{e}sentons, en d\'{e}tail, une dualit%
\'{e} d'importance capitale entre la th\'{e}orie-M sur $K3$ et la corde h%
\'{e}t\'{e}rotique sur $T^{3}$ \`{a} sept dimensions. Alors que dans la
seconde partie, nous d\'{e}voilons certaines des caract\'{e}ristiques et
propri\'{e}t\'{e}s de la vari\'{e}t\'{e} compacte qui nous permet d'aboutir 
\`{a} un mod\`{e}le \`{a} quatre dimensions avec un nombre minimal de
supercharges. Cette vari\'{e}t\'{e} n'est autre qu'une vari\'{e}t\'{e} \`{a}
sept dimensions ayant un groupe d'holonomie ${\bf G}_{2}$ que nous notons $%
X. $

La quatri\`{e}me section est \'{e}galement constitu\'{e}e de deux parties.

Au cours de la premi\`{e}re, nous pr\'{e}sentons une r\'{e}alisation de la
vari\'{e}t\'{e} $X$ dont les singularit\'{e}s garantissent l'obtention d'un
mod\`{e}le physique \`{a} $D=4$ avec ${\cal N}=1$ ayant un groupe de jauge
non ab\'{e}lien. Or, inspir\'{e}e de la dualit\'{e} corde-corde la plus int%
\'{e}ressante existant \`{a} six dimensions entre la supercorde type IIA sur 
$K3,$ avec des singularit\'{e}s $ADE,$ et la supercorde h\'{e}t\'{e}rotique
sur $T^{4}$, nous d\'{e}duisons que cette r\'{e}alisation ne peut \^{e}tre
qu'une fibration $K3$ sur une base \`{a} trois dimensions et dont le premier
nombre de Betti est nul.

Dans la deuxi\`{e}me partie de cette section, nous nous sommes bas\'{e}es
sur un r\'{e}sultat des travaux de Katz et Vafa \cite{ref20} concernant l'ing%
\'{e}nierie g\'{e}om\'{e}trique de la mati\`{e}re charg\'{e}e pour type IIA
sur une Calabi-Yau de dimension trois complexes. Ce travail nous a permis de
compl\'{e}ter la construction de $X$ afin qu'elle fournisse de la mati\`{e}%
re chirale en plus des sym\'{e}tries non ab\'{e}liennes dans le spectre du
mod\`{e}le r\'{e}sultant \`{a} quatre dimensions.

Enfin, nous terminons par une conclusion.

\section{Th\'{e}orie-M \`{a} partir des Supercordes}

\subsection{Th\'{e}orie des cordes et des supercordes}

A la fin des ann\'{e}es 1960, la th\'{e}orie des cordes (ou la th\'{e}orie
des cordes bosoniques) a surgit afin de d\'{e}crire la force nucl\'{e}aire
forte. Quelques ann\'{e}es plus tard, en 1971, l'inclusion des fermions a
entra\^{\i}n\'{e} l'\'{e}tude de la corde supersym\'{e}trique appel\'{e}e
aussi supercorde. Cependant, le d\'{e}veloppement rapide qu'a connu la
Chromodynamique quantique (QCD) en 1973 l'a incit\'{e} \`{a} \^{e}tre
reconnue comme th\'{e}orie capable de d\'{e}crire les interactions fortes.
Ainsi la th\'{e}orie des cordes (supercordes) fut priv\'{e}e de son but
initial, mais elle acquiert un autre objectif puisqu'elle fut estim\'{e}e
par la suite d'\^{e}tre celle qui constituera un jour ''une th\'{e}orie du
tout fondamentale''.\newline
Effectivement, il existe parmi ses \'{e}tats non massifs, un \'{e}tat qui a
un spin 2. En 1974, Sherk et Shwartz et ind\'{e}pendamment Yoneya ont montr%
\'{e} que cette particule interagit comme un graviton \cite{ref0,ref1}. Ce
fait a permis \`{a} la th\'{e}orie d'inclure la relativit\'{e} g\'{e}n\'{e}%
rale. Par cons\'{e}quent, les trois physiciens ont propos\'{e} d'une part
que la th\'{e}orie doit \^{e}tre utilis\'{e}e pour l'unification, et d'autre
part que l'\'{e}chelle de longueur de la corde doit \^{e}tre comparable \`{a}
la longueur de Plank.

\subsubsection{Th\'{e}orie des cordes}

Contrairement \`{a} la th\'{e}orie quantique des champs o\`{u} les objets
fondamentaux sont consid\'{e}r\'{e}s comme des particules ponctuelles de
dimension nulle, la th\'{e}orie des cordes les voit plut\^{o}t comme des
objets \'{e}tendus \cite{ref1} de dimension un poss\'{e}dant une tension $T$
. De plus, certains de ces objets (cordes) sont ferm\'{e}s vus comme des
boucles, d'autres sont ouverts assimil\'{e}s \`{a} des petits segments avec
des conditions aux bords de Dirichlet ou de Neumann \cite{ref2}.\newline
D'autre part, puisque la th\'{e}orie des cordes est une th\'{e}orie
quantique relativiste qui inclut la gravit\'{e}, alors elle doit susciter
l'existence d'un lien entre les constantes fondamentales. Effectivement, la
vitesse de la lumi\`{e}re $c$, $h$ la constante de Planck et $G$ la
constante gravitationnelle de Newton sont tous reli\'{e}es dans les formules
suivantes.

\begin{itemize}
\item L'\'{e}chelle de Planck: 
\begin{eqnarray}
l_{p} &=&\left( \frac{hG}{2\pi c^{3}}\right) ^{\frac{3}{2}}  \nonumber \\
&=&1.6\times 10^{-33} cm
\end{eqnarray}

\item La masse de Planck: 
\begin{eqnarray}
m_{p} &=&\left( \frac{hc}{2\pi G}\right) ^{\frac{1}{2}}  \nonumber \\
&=&1.2\times 10^{19}{GeV}/{{c}^{2}}.
\end{eqnarray}
\end{itemize}

Lors de son mouvement, la corde bosonique (soit ferm\'{e}e ou bien ouverte)
balaye une surface bidimensionnelle appel\'{e}e surface d'univers. De cette
propri\'{e}t\'{e}, il d\'{e}coule que la th\'{e}orie classique des cordes
peut \^{e}tre consid\'{e}r\'{e}e comme une th\'{e}orie des champs
bidimensionnelle conforme. Par suite, au niveau quantique, les contraintes
d'invariance conforme exigent que la dimension de l'espace-temps soit $D=26$
au lieu de $(1+3)$ dimensions habituelles. Bien que cette th\'{e}orie des
cordes bosoniques poss\`{e}de d'importantes caract\'{e}ristiques, elle n'est
pas consistante en raison de:

\begin{enumerate}
\item La dimension critique $D=26$.

\item L'absence des fermions n\'{e}cessaires pour d\'{e}crire la mati\`{e}re.

\item La pr\'{e}sence du tachyon dans son spectre (particule de masse au carr%
\'{e} n\'{e}gative).
\end{enumerate}

\subsubsection{Th\'{e}orie des supercordes}

Suite \`{a} tout ces probl\`{e}mes fut introduite la th\'{e}orie des
supercordes. Elle est consid\'{e}r\'{e}e comme une g\'{e}n\'{e}ralisation
supersym\'{e}trique du mod\`{e}le de la corde bosonique, et ceci en ajoutant
des champs fermioniques sur la surface d'univers. Cette augmentation par des
spineurs donne lieu \`{a} une sym\'{e}trie plus riche \`{a} savoir la sym%
\'{e}trie superconforme. De plus l'adjonction des champs fermioniques \`{a}
la corde bosonique, qui sont \`{a} la fois des spineurs de Majorana \`{a}
deux dimensions et des vecteurs de Lorentz, a pour cons\'{e}quence la r\'{e}%
duction de la dimension critique de 26 \`{a} 10 dimensions. Dans cette th%
\'{e}orie, le probl\`{e}me du tachyon fut surmont\'{e} par la projection de
Gliozzi, Sherk et Olive (GSO) \cite{ref2,ref3} qui consiste \`{a} supprimer
quelques \'{e}tats de la th\'{e}orie et par ailleurs, permet de se limiter 
\`{a} un sous espace des \'{e}tats du spectre o\`{u} l'existence d'un nombre 
\'{e}gal de particules bosoniques et fermioniques \`{a} chaque niveau
d'excitation est garantit. Ce spectre des \'{e}tats non massifs de la
supercorde aussi bien que de la corde bosonique contient un \'{e}tat non
massif de spin 2 qui peut \^{e}tre identifi\'{e} au m\'{e}diateur des int%
\'{e}ractions gravitationnelles: le graviton.

Entre 1984-1985, apparu ''la premi\`{e}re r\'{e}volution de la supercorde''
qui a impliqu\'{e} l'existence de 5 types de supercordes \`{a} D=10 class%
\'{e}es comme suit:

({\bf a})- Les supercordes ayant une supersym\'{e}trie d'espace-temps ${\cal %
N}=1$ \`{a} dix dimensions comprenant:

\begin{enumerate}
\item La supercorde de type I avec un groupe de jauge $SO(32)$.

\item La supercorde h\'{e}t\'{e}rotique $SO(32)$.

\item La supercorde h\'{e}t\'{e}rotique $E_{8}\times E_{8}$.\newline
\end{enumerate}

({\bf b})- Les supercordes ayant une supersym\'{e}trie d'espace-temps ${\cal %
N}=2$:

\begin{enumerate}
\item La supercorde non chirale type IIA.

\item La supercorde chirale type IIB.
\end{enumerate}

\subsubsection{Spectre des supercordes}

Le spectre des \'{e}tats non massifs des cinq mod\`{e}les contient le
dilaton $\phi $, dont $g_{s}=e^{\phi }$ est la constante de couplage de la th%
\'{e}orie, le graviton de spin 2 et des tenseurs antisym\'{e}triques de
jauge g\'{e}n\'{e}ralisant la notion du potentiel vecteur $A_{\mu }$ \`{a}
des tenseurs antisym\'{e}triques \cite{ref3} $A_{\mu _{1}\ldots \mu _{p+1}}$ 
\`{a} $p+1$ indices (($p+1$)-formes, $p=1,2,\ldots $). Plus pr\'{e}cis\'{e}%
ment nous avons pour:

\begin{itemize}
\item {\it {\underline{ Les th\'{e}ories type II}}}
\end{itemize}

Le spectre totale de ces deux mod\`{e}les est donn\'{e} par:\newline
{\bf i) Secteur bosonique:} Ce secteur se scinde en deux types.\newline
($\alpha $) Bosons {\bf NS-NS}: Ces bosons sont les m\^{e}mes autant pour
type IIA que pour type IIB. Ils sont r\'{e}partis en un dilaton $\phi $, un
graviton $g_{\mu \nu }$ et un tenseur antisym\'{e}trique $B_{\mu \nu }$.%
\newline
($\beta $) Bosons {\bf R-R}: Ceux-ci par contre d\'{e}pendent du choix de
chiralit\'{e} relative des spineurs. \newline
Dans le cas de la th\'{e}orie IIA qui est non chirale, nous trouvons un
vecteur de jauge $A_{\mu }$ et un tenseur 3-forme $C_{\mu \nu \rho }$ antisym%
\'{e}trique. Tandis que dans le cas de la th\'{e}orie type IIB chirale, les
bosons {\bf R-R} sont form\'{e}s d'un champ scalaire $\chi $ (axion), une
2-forme $\tilde{B}_{\mu \nu }$ antisym\'{e}trique et une 4-forme $D_{\mu \nu
\rho \sigma } $ antisym\'{e}trique auto-duale.\newline
{\bf ii) Secteur fermionique:} Les deux secteurs fermioniques {\bf R-NS} et 
{\bf NS-R} sont identiques et contiennent un fermion et un gravitino.

\begin{itemize}
\item {\it {\underline{ La supercorde h\'{e}t\'{e}rotique:}}}
\end{itemize}

Dans ce type par contre le spectre bosonique est donn\'{e} par: 
\begin{eqnarray}
(g_{\mu \nu },B_{\mu \nu },\phi ),  \nonumber \\
A_{\mu }=A_{\mu }^{a}T_{a},
\end{eqnarray}
avec $\quad a=1,\ldots ,\dim SO(32) $ ou $\;\dim E_{8}\times E_{8}.$ Alors
que la partie fermionique est constitu\'{e}e du gravitino et du jaugino,
partenaires supersym\'{e}triques du graviton et du champ de jauge
respectivement dans la repr\'{e}sentation adjointe du groupe de jauge.

\begin{itemize}
\item {\it {\underline{ La th\'{e}orie type I}}}
\end{itemize}

Son spectre est obtenu \`{a} partir des supercordes type IIB. Il correspond
au dilaton $\phi $, graviton $g_{\mu \nu }$ du secteur de {\bf NS-NS} de la
th\'{e}orie de type IIB, le tenseur antisym\'{e}trique $\tilde{B}_{\mu \nu }$
du secteur {\bf R-R} des supercordes ferm\'{e}es, les champs de jauge $%
SO(32) $ du secteur des cordes ouvertes et leurs partenaires fermioniques
sous la supersym\'{e}trie ${\cal N}=1$ \`{a} dix dimensions. La partie
bosonique de ce mod\`{e}le est donn\'{e}e par 
\begin{equation}
(g_{\mu \nu },\tilde{B}_{\mu \nu },\phi ),A_{\mu }=A_{\mu }^{a}T_{a},
\end{equation}
avec $\quad a=1,\ldots \dim SO(32)$. Nous notons que le spectre de la th\'{e}%
orie type I est identique \`{a} celui de la th\'{e}orie h\'{e}t\'{e}rotique
de groupe de jauge $SO(32)$.

\subsection{Compactification en th\'{e}orie des supercordes}

Afin que la th\'{e}orie des supercordes d\'{e}crive notre univers qui est
seulement \`{a} $(3+1)$ dimensions non compactes, il s'est av\'{e}r\'{e} n%
\'{e}cessaire d'utiliser la m\'{e}thode de compactification \cite{ref5}. Ce
proc\'{e}d\'{e} qui a vu le jour suite aux travaux successifs de Kaluza et
Klein, suppose que certaines des dix dimensions sont compactes et non
observables \`{a} notre \'{e}chelle. De ce fait, il faudrait consid\'{e}rer
des g\'{e}om\'{e}tries o\`{u} l'espace de Minkowski \`{a} 10 dimensions $%
M_{10}$ se d\'{e}compose en:

\begin{enumerate}
\item Une vari\'{e}t\'{e} non compacte correspondant \`{a} l'espace-temps de
Minkowski usuel $M_{4}$.

\item Une vari\'{e}t\'{e} compacte $K_{6}$ de dimension 6 et de volume tr%
\`{e}s petit devant notre \'{e}chelle d'observation,
\end{enumerate}

\begin{equation}
M_{10}\rightarrow {M_{4}}\times {K_{6}}.
\end{equation}
Ce sc\'{e}nario s'\'{e}tend m\^{e}me \`{a} des compactifications sur des vari%
\'{e}t\'{e}s compactes $K_{d}$ mais vers des espaces-temps arbitraires $%
M_{10-d}$. Cependant puisque un spineur \`{a} dix dimensions se d\'{e}%
compose en un spineur sur l'espace $K_{d}$ et un spineur \`{a} $(10-d)$, il
en r\'{e}sulte que le nombre de supersym\'{e}tries pr\'{e}serv\'{e}es \`{a} $%
(10-d)$ dimensions de l'espace-temps est \'{e}gal au nombre de spineurs
covariantiquement constants sur $K_{d}$. Par cons\'{e}quent, ces espaces
compactes $K_{d}$ sont choisis de sorte qu'ils pr\'{e}servent un certain
nombre de supersym\'{e}tries des supercordes qui se propagent dans un
espace-temps de dimension $(10-d)$. Par suite, les vari\'{e}t\'{e}s $K_{d} $
sont class\'{e}es selon les diff\'{e}rents mod\`{e}les de supercordes r\'{e}%
sultants \`{a} $(10-d)$ dimensions \cite{ref1}.\newline
Effectivement, parmi les multiples compactifications qui existent, nous
citons trois des exemples les plus utilis\'{e}s en th\'{e}orie des
supercordes:\newline
$\bullet $ La compactification toro\"{\i}dale sur un tore $T^{d}$ ayant un
groupe d'holonomie trivial $U(1)^{d}$. Celle-ci conduit \`{a} un mod\`{e}le
avec 32 supercharges.\newline
$\bullet $ La compactification sur l'hypersurface $K3$ de groupe d'holonomie 
$SU(2)$ pr\'{e}servant la moiti\'{e} des supercharges initiales.\newline
$\bullet $ La compactifiaction sur des vari\'{e}t\'{e}s de Calabi-Yau \`{a}
six dimensions $X_{3}$ ayant un groupe d'holonomie $SU(3)$.\newline
Ainsi, il est clair \`{a} pr\'{e}sent que la compactification nous permet
non seulement de d\'{e}crire les diff\'{e}rents mod\`{e}les de supercordes 
\`{a} des dimensions inf\'{e}rieurs mais aussi de r\'{e}duire leurs nombres
de charges supersym\'{e}triques.

D'autre part, la compactification des cinq mod\`{e}les de supercordes d\'{e}j%
\`{a} cit\'{e}s donne plusieurs th\'{e}ories dans les dimensions inf\'{e}%
rieures. Chacune de ces th\'{e}ories est param\'{e}tris\'{e}e par les
modules suivants:

\begin{enumerate}
\item La constante de couplage de la supercorde $g_{s}=e^{\phi }$, o\`{u} $%
\phi $ est la valeur moyenne du dilaton dans le vide.

\item Les modules g\'{e}om\'{e}triques de la vari\'{e}t\'{e} complexe
compacte dont le nombre provient des diff\'{e}rents choix possibles de la m%
\'{e}trique. Ce nombre est donn\'{e} par les d\'{e}formations de Kahler et
les d\'{e}formations de la structure complexe.

\item Les valeurs moyennes des champs antisym\'{e}triques des secteurs {\bf %
NS-NS} et {\bf R-R} et des champs de jauge.
\end{enumerate}

Ces trois types de valeurs moyennes param\'{e}trisent l'espace des modules
de la th\'{e}orie compactifi\'{e}e. Dans la r\'{e}gion de l'espace des
modules o\`{u} la constante de couplage est faible, la th\'{e}orie
perturbative est rev\'{e}lante. Alors que dans la r\'{e}gion o\`{u} la
constante de couplage est forte, c'est plut\^{o}t le r\'{e}gime non
perturbatif qui est dominant.

\subsection{Dualit\'{e} en th\'{e}orie des supercordes}

En d\'{e}pit de ses succ\`{e}s majeurs, la th\'{e}orie des supercordes conna%
\^{\i}t des doutes persistants puisqu'au lieu d'unifier les th\'{e}ories,
nous nous trouvons devant cinq types de mod\`{e}les consistants \`{a} dix
dimensions. Par suite apr\`{e}s compactification, un grand nombre de th\'{e}%
orie des supercordes surgit aux dimensions inf\'{e}rieures. Heureusement,
l'ann\'{e}e 1995 donna naissance \`{a} ''la seconde r\'{e}volution de la th%
\'{e}orie des supercordes'' gr\^{a}ce \`{a} laquelle l'int\'{e}r\^{e}t fut
apport\'{e} \`{a} l'\'{e}tude des sym\'{e}tries non perturbatives de ces th%
\'{e}ories. Dans ce contexte, les cinq mod\`{e}les qui apparaissaient diff%
\'{e}rents dans leurs descriptions perturbatives (\`{a} faible couplage),
sont en fait reli\'{e}s \`{a} couplage fort par les diff\'{e}rentes dualit%
\'{e}s des cordes.\newline
Rappelons que ce concept de dualit\'{e} d\'{e}j\`{a} connu en m\'{e}canique
quantique entre onde et corpuscule, signifiant l'existence de deux approches
diff\'{e}rentes du m\^{e}me syst\`{e}me physique, joue un r\^{o}le important
dans la th\'{e}orie des supercordes en permettant la connection entre ses
diff\'{e}rents types \cite{ref1,ref4,ref5}.\newline
Voyons bri\`{e}vement des exemples de sym\`{e}trie de dualit\'{e}.

\subsubsection{Dualit\'{e}-T}

Cette dualit\'{e} poss\'{e}de une cons\'{e}quence int\'{e}ressante dans le
cas des th\'{e}ories type II. En effet consid\'{e}rons une th\'{e}orie type
II et compactifions la neuvi\`{e}me direction sur un cercle de rayon R.
Alors la limite $R\rightarrow 0$ est \'{e}quivalente \`{a} la limite $%
R^{\prime }\rightarrow \infty $ pour la coordonn\'{e}e duale avec $R^{\prime
}={\frac{\alpha ^{\prime }}{R}}$. Cette sym\'{e}trie signifie que la
compactification sur un cercle de petit rayon est \'{e}quivalente \`{a} la
compactification sur un cercle de grand rayon et que les limites $%
R\rightarrow 0$ et $R^{\prime }\rightarrow \infty $ sont physiquement
identiques. De plus, cette dualit\'{e}-T change la chiralit\'{e} des \'{e}%
tats \`{a} mouvement gauche, et par cons\'{e}quent transforme la th\'{e}orie
type IIA en type IIB et vice-versa. \newline
Elle rend m\^{e}me les th\'{e}ories h\'{e}t\'{e}rotiques $SO(32)$ et $%
E_{8}\times E_{8}$ \'{e}quivalentes \`{a} neuf dimensions. \newline
Ainsi la dualit\'{e}-T qui est une sym\'{e}trie perturbative r\'{e}sultante
d'une compactification \`{a} des dimensions inf\'{e}rieures, relie deux th%
\'{e}ories diff\'{e}rentes dans la m\^{e}me r\'{e}gion de faible couplage.

\subsubsection{Dualit\'{e}-S}

Cette dualit\'{e} par contre relie deux r\'{e}gimes de couplages diff\'{e}%
rents. Elle est donn\'{e}e par l'inversion du couplage $g_{s}$, ce qui
signifie que 
\begin{equation}
g_{s}\longrightarrow {\frac{1}{g_{s}}}.
\end{equation}
Cette sym\'{e}trie de dualit\'{e} nous permet de d\'{e}crire le couplage
fort (faible) d'une th\'{e}orie \`{a} l'aide du r\'{e}gime \`{a} couplage
faible (fort) de sa th\'{e}orie duale.\newline
A dix dimensions, la th\'{e}orie h\'{e}t\'{e}rotique $SO(32)$ et la th\'{e}%
orie type I $SO(32)$ sont \'{e}quivalentes par inversement de couplage de la
corde $g_{het}\leftrightarrow {\frac{1}{g_{typeI}}}$. D'autre part la dualit%
\'{e}-S permet l'\'{e}change de la constante de couplage $%
g_{IIB}\leftrightarrow {\frac{1}{g_{IIB}}}$ de la th\'{e}orie type IIB qui
est dite auto-duale sous la sym\'{e}trie $Sl(2,{\bf Z})$.\newline

En conclusion, les deux th\'{e}ories ${\cal N}=2$ et les trois th\'{e}ories $%
{\cal N}=1$ sont s\'{e}par\'{e}ment conn\'{e}ct\'{e}es par le biais de ces
deux dualit\'{e}s. Mais En pratique\cite{ref1,ref4}, il se trouve que le cas
le plus important de dualit\'{e} corde-corde est celui reliant la supercorde
IIA sur une vari\'{e}t\'{e} $K3$ et la supercorde h\'{e}t\'{e}rotique sur un
tore $T^{4}$ . Ces deux mod\`{e}les pr\'{e}sentent le m\^{e}me espace des
modules 
\begin{equation}
{\bf R^{+}}\times {\frac{SO(20,4,{\bf R})}{SO(20)\times SO(4)}}
\end{equation}
et la sym\'{e}trie de jauge perturbative des supercordes h\'{e}t\'{e}%
rotiques sur $T^{4}$ est identifi\'{e}e avec les singularit\'{e}s de $K3.$
Ainsi cette dualit\'{e} permet de relier les th\'{e}ories ${\cal N}=1$ et $%
{\cal N}=2$ \`{a} six dimensions.

\subsection{Th\'{e}orie-M}

\subsubsection{Introduction}

L'une des cons\'{e}quences majeures de l'\'{e}tude de la dualit\'{e} entre
les cinq mod\`{e}les de supercordes perturbatives \`{a} dix dimensions, est
le fait qu'on peut les voir comme des limites non perturbatives d'une seule
th\'{e}orie appel\'{e}e {\bf th\'{e}orie-M }\`{a} onze dimensions.\newline
La lettre '' M '' a plusieurs interpr\'{e}tations parmi lesquelles:

\begin{description}
\item[-Magique ou Myst\'{e}rieuse.] 

\item[-Membrane] puisqu'elle contient M2-brane.

\item[-Mother] comme m\`{e}re de toute les th\'{e}ories.

\item[-Matrice] comme une autre approche possible de la th\'{e}orie.
\end{description}

Selon Witten, cette th\'{e}orie-M est d\'{e}crite \`{a} faible \'{e}nergie
par une th\'{e}orie de supergravit\'{e} \`{a} onze dimensions ayant 32
supercharges conserv\'{e}es (${\cal N}=1$).\newline
Construite en 1978 par Cremmer, Julia et Sherk la supergravit\'{e} poss\`{e}%
de 3 types de champs:

\begin{enumerate}
\item Le graviton $g_{MN}$ ( avec 44 polarisations ).

\item Le tenseur de jauge ( 3-forme ) $C_{MNP}$ ( avec 84 polarisations ).

\item Le gravitino $\Psi_{M}^{a}$ ( avec 128 polarisations ).
\end{enumerate}

Ces champs constituent le {\bf "supermultiplet gravitationnel"}. En plus de
ces champs, le spectre de la th\'{e}orie-M (supergravit\'{e} \`{a} faible 
\'{e}nergie ) contient \'{e}galement des objets solitoniques nomm\'{e}s
M2-branes qui se couplent \`{a} $C_{MNP}$, et leurs duals magn\'{e}tiques
M5-branes. \newline
Cette th\'{e}orie supersym\'{e}trique qui est g\'{e}n\'{e}ralement
covariante poss\`{e}de une invariance de jauge sous laquelle: 
\begin{equation}
\delta C = d\lambda
\end{equation}
avec $\lambda $ est une 2-forme. Ainsi le champ invariant de jauge est la d%
\'{e}riv\'{e}e de $C$ que nous notons $F$.\newline
L'action des champs bosoniques est donn\'{e}e par 
\begin{equation}
S=\int{{\sqrt{g}R} -{\frac{1}{2}F\wedge{\ast F}}-{\frac{1}{6}C\wedge F
\wedge F}}.
\end{equation}
Par cons\'{e}quent les \'{e}quations de mouvement de $C$ et $g$ prennent les
formes 
\begin{equation}
d\ast{F}=F\wedge{F}
\end{equation}
et 
\begin{equation}
R_{MN}=T_{MN}\left( C\right)
\end{equation}
o\`{u} $T$ est le tenseur \'{e}nergie-impulsion du champ $C$.

\subsubsection{Th\'{e}orie-M et th\'{e}orie des supercordes}

(a) {\it {Th\'{e}orie-M et th\'{e}orie type IIA }}

Contrairement \`{a} type IIB, la th\'{e}orie type IIA n'a pas de sym\'{e}%
trie reliant les couplages fort et faible. Il a \'{e}t\'{e} conjectur\'{e}
que la limite \`{a} couplage fort de la th\'{e}orie type IIA n'est pas une th%
\'{e}orie \`{a} dix dimensions, mais plut\^{o}t une th\'{e}orie \`{a} onze
dimensions qui est la th\'{e}orie-M \cite{ref6}. Pour expliquer cette
relation existante entre ces deux th\'{e}ories, nous identifions le spectre
qui caract\'{e}rise chacune d'elles \`{a} dix dimensions.\newline
La compactification de la th\'{e}orie-M sur un cercle de rayon $R_{10}$ qui
est donn\'{e} par: 
\begin{equation}
R_{10}=l_{11}g_{s}^{\frac{2}{3}}
\end{equation}
avec $g_{s}$ est la constante de couplage de la th\'{e}orie type IIA et $%
l_{11}=\alpha ^{\prime }{g_{s}^{\frac{1}{3}}}$ est l'echelle de Plank \`{a}
onze dimensions. Ainsi le rayon $R_{10}$ de la onzi\`{e}me dimension est
bien reli\'{e} \`{a} $g_{s}$ de type IIA.\newline
De plus pour $M,N,P=0,.....,10$ et $\mu ,\nu ,\rho =0,....,9$, cette
compactification se r\'{e}sume comme suit:

\begin{description}
\item[-La m\'{e}trique] \`{a} onze dimensions $g_{MN}$ se r\'{e}duit en:

\begin{enumerate}
\item La m\'{e}trique $g_{\mu\nu}$ \`{a} dix dimensions.

\item Le champ de jauge $g_{\mu10}=A_{\mu}$.

\item Le dilaton $\phi = g_{1010}$.
\end{enumerate}
\end{description}

En terme de nombre de degr\'{e} de lib\'{e}rt\'{e} ceci s'exprime par 
\begin{equation}
44=35\oplus8\oplus1.
\end{equation}

\begin{description}
\item[-Le tenseur de jauge $C_{MNP}$] donne \`{a} son tour par la r\'{e}%
duction dimensionnelle:

\begin{enumerate}
\item La 3-forme $C_{\mu\nu\rho}.$

\item La 2-forme $B_{\mu\nu}=C_{\mu\nu10},$
\end{enumerate}
\end{description}

D'o\`{u} nous pouvons l'\'{e}crire comme suit: 
\begin{equation}
84=56 \oplus28.
\end{equation}
Par cons\'{e}quent, ces champs co\"{\i}ncident avec les champs bosoniques de
la th\'{e}orie type IIA dans ses deux secteurs {\bf R-R} et {\bf NS-NS}.
Autrement dit, \`{a} faible couplage $g_{s}\longrightarrow0$ et $%
(R_{10}\longrightarrow0 )$ nous obtenons la th\'{e}orie type IIA \`{a} dix
dimensions. Tandis qu'\`{a} fort couplage $g_{s}\longrightarrow\infty$ le
rayon de compactification devient infini ($R_{10}\longrightarrow\infty$),
alors nous retrouvons une th\'{e}orie dont la onzi\`{e}me dimension est non
compacte et qui n'est autre que la th\'{e}orie-M.\newline
D'autre part, puisque le spectre non massif de la th\'{e}orie-M contient
aussi des objets non perturbatifs, alors il faut s'attendre \`{a} retrouver
les branes de la th\'{e}orie type IIA apr\`{e}s la r\'{e}duction
dimensionnelle sur un cercle vers dix dimensions. \newline
En effet,\newline
$\diamond$ la M-2 brane transverse \`{a} la direction compacte est l'origine
de D-2 brane \`{a} D=10. \newline
$\diamond$ la M-2 brane envelopp\'{e}e autour de la dimension compacte donne
la corde fondamentale.\newline
$\diamond$ La M-5 brane transverse \`{a} la onzi\`{e}me dimension donne lieu 
\`{a} NS-5 brane. \newline
$\diamond$ La M-5 brane envelopp\'{e}e autour de la dimension compacte
produit la D-4 brane.\newline
$\diamond$ Par contre la D-0 brane (duale de D-6 brane) est coupl\'{e}e \`{a}
$g_{\mu10}.$ \newline
Ainsi tous les objets perturbatifs et non perturbatifs de type IIA sont
obtenus \`{a} partir de la th\'{e}orie-M.

(b){\it {\ Th\'{e}orie-M et corde h\'{e}t\'{e}rotique $E_{8}\times E_{8}$}}

Une autre approche pour mieux comprendre la th\'{e}orie-M est de consid\'{e}%
rer son comportement sur une orbifold particuli\`{e}re \`{a} onze dimension $%
\frac{{\bf S^{1}}}{{\bf Z_{2}}}$, o\`{u} ${\bf Z_{2}}$ agit sur ${\bf S^{1}}$
comme inversion de $x^{11}\longrightarrow -x^{11}.$ L'espace r\'{e}sultant
est vu comme un segment de points fixes $0$ et $\pi .$ \newline
La th\'{e}orie-M sur ${\bf R}^{10}\times {\frac{{\bf S^{1}}}{{\bf Z_{2}}}}$
se r\'{e}duit dans le cas de basse \'{e}nergie \`{a} une th\'{e}orie de
supergravit\'{e} \`{a} $D=10$ avec ${\cal N}=1$. Or, il y a trois th\'{e}%
ories de supercordes poss\'{e}dant cette structure \`{a} basse \'{e}nergie,

\begin{itemize}
\item H\'{e}t\'{e}rotique $E_{8}\times E_{8}$.

\item H\'{e}t\'{e}rotique et type I ayant chacune $SO(32)$ comme groupe de
jauge.
\end{itemize}

Ainsi, si le rayon de $S^{1}$ tend vers $0$, la th\'{e}orie-M sur ${\bf R}%
^{10}\times {\frac{{\bf S^{1}}}{{\bf Z_{2}}}}$ se r\'{e}duit s\^{u}rement 
\`{a} l'une de ces trois th\'{e}ories.\newline
Selon Horava et Witten \cite{ref6}, cette th\'{e}orie ne peut \^{e}tre que
la corde h\'{e}t\'{e}rotique $E_{8}\times E_{8}$. Puisque la construction de
cet orbifold entra\^{\i}ne g\'{e}n\'{e}riquement l'existence d'\'{e}tats
twist\'{e}s localis\'{e}s sur les deux points du segment. Alors physiquement
ceci peut \^{e}tre int\'{e}rpr\'{e}t\'{e} comme l'existence des D9-branes 
\`{a} chaque extr\'{e}mit\'{e} du segment. Ainsi elles portent des degr\'{e}
de libert\'{e} de jauge qui nous indiquent que les groupes de jauge
possibles \`{a} $D=10$ ${\cal N}=1$ doivent avoir une dimension $496$. Et du
fait que nous avons deux hyperplans fix\'{e}s \`{a} $x=0$ et $x=\pi$ donc la
dimension est exactement 248 qui n'est autre que celle de $E_{8}$. \newline
Finalement, la th\'{e}orie-M sur l'orbifold d\'{e}j\`{a} cit\'{e}e est reli%
\'{e}e \`{a} la corde h\'{e}t\'{e}rotique $E_{8}\times E_{8}$ avec un $E_{8}$
se propageant en chacun des deux hyperplans qui forment les bords de
l'espace-temps.\newline

En conclusion, si la dualit\'{e}-S nous a permis de d\'{e}terminer la limite
de couplage-fort des mod\`{e}les type IIB, type I et h\'{e}t\'{e}rotique $%
SO(32)$ et bien seule l'introduction de la th\'{e}orie-M, qui est \`{a} $%
D=11 $, est capable de d\'{e}terminer la limite de couplage-fort des mod\`{e}%
les type IIA et h\'{e}t\'{e}rotique $E_{8}\times E_{8}$. Ainsi cette th\'{e}%
orie permet de faire appel \`{a} la physique non perturbative des cinq th%
\'{e}ories des supercordes dans laquelle, les cinq mod\`{e}les qui
apparaissaient diff\'{e}rents dans leurs descriptions perturbatives
faiblement coupl\'{e}es, sont en fait reli\'{e}s \`{a} couplage fort par le
proc\'{e}de de dualit\'{e}.

\section{Compactification de La Th\'{e}orie-M}

Dans la section pr\'{e}c\'{e}dente, nous avons constat\'{e} que les sym\'{e}%
tries de dualit\'{e} ont conduit \`{a} une r\'{e}volution dans la conception
des th\'{e}ories des supercordes, puisqu'elles nous ont permis de voir les
cinq mod\`{e}les des supercordes comme des manifestations d'une seule th\'{e}%
orie qui est la th\'{e}orie-M. Or \'{e}tant donn\'{e} que cette derni\`{e}re
est d\'{e}crite \`{a} faible \'{e}nergie par une th\'{e}orie de supergravit%
\'{e} \`{a} onze dimensions. Alors il sera int\'{e}ressant d'essayer de la
ramener \`{a} une th\'{e}orie ayant un nombre minimal de supercharges et se
propageant au monde r\'{e}el \`{a} $3+1$ dimensions habituelles. De ce fait,
nous avons besoin de compactifier sur une vari\'{e}t\'{e} de dimension sept.
Mais avant d'aborder cette r\'{e}duction dimensionnelle qui n\'{e}cessite
l'introduction d'une vari\'{e}t\'{e} de groupe d'holonomie ${\bf G}_{2},$
nous allons en premier lieu utiliser l'un des r\'{e}sultats remarquables
offerts par la dualit\'{e}\cite{ref1} reliant la supercorde IIA sur une vari%
\'{e}t\'{e} $K3$ et la supercorde h\'{e}t\'{e}rotique sur un tore $T^{4}$ .
Ce r\'{e}sultat n'est autre que la relation entre la th\'{e}orie-M et la
corde h\'{e}t\'{e}rotique \`{a} sept dimensions.

\subsection{Dualit\'{e} entre th\'{e}orie-M et corde h\'{e}t\'{e}rotique 
\`{a} sept dimensions}

La th\'{e}orie-M compactifi\'{e}e sur $K3$ est \'{e}quivalente \cite{ref7} 
\`{a} la corde h\'{e}t\'{e}rotique sur un tore $T^{3}$ . Afin d'expliciter
ce r\'{e}sultat, commen\c{c}ons tout d'abord par rappeler que $K3$ est une
vari\'{e}t\'{e} Kahl\'{e}rienne compacte de dimension r\'{e}elle 4 et de
courbure de Ricci nulle \cite{ref8}. Elle a pour groupe d'holonomie le
groupe $SU(2)$ qui assure l'existence de spineurs covariantiquement
constants. De plus la compactification sur $K3$ pr\'{e}serve la moiti\'{e}
des charges supersym\'{e}triques initiales. Localement, cette vari\'{e}t\'{e}
a pour espace des modules: 
\begin{equation}
{\bf {\cal M}}(K3)={\bf R}^{+}\times \frac{SO(3,19)}{SO(3)\times SO(19)}
\label{e1}
\end{equation}
qui contient 58 param\`{e}tres r\'{e}els et o\`{u} ${\bf R}^{+}$ correspond 
\`{a} un facteur d'\'{e}chelle global de la m\'{e}trique de $K3$, qui peut 
\^{e}tre vu comme le volume de $K3$.

Dans un point r\'{e}gulier de ${\bf {\cal M}}(K3)$, les modules g\'{e}om\'{e}%
triques de $K3$ suffisent pour d\'{e}crire la compactification de la th\'{e}%
orie-M sur cet espace. Effectivement \`{a} sept dimensions, nous retrouvons
exactement 58 champs scalaires non massifs qui ne sont autres que les
fluctuations de la m\'{e}trique sur $K3$, surtout que la 3-forme ne g\'{e}n%
\`{e}re aucun degr\'{e} de libert\'{e} suppl\'{e}mentaire sur $K3$ puisque $%
b_{3}(K3)=0$. D'autre part, en plus de la m\'{e}trique et la 3-forme, le
groupe $H^{2}(K3,{\bf R})={\bf R^{22}}$ indique l'existence de 22 classes lin%
\'{e}airement ind\'{e}pendantes des 2-formes harmoniques. Ainsi nous avons
un groupe de jauge ${U(1)}^{22}$ caract\'{e}risant les 22 champs de jauge r%
\'{e}sultants \`{a} sept dimensions. Et puisque $b_{1}(K3)=0 $, ce fait
exige que le spectre donn\'{e} est le spectre complet du mod\`{e}le obtenu 
\`{a} sept dimensions ayant 16 supercharges.

Reste maintenant \`{a} prouver qu'il s'agit exactement du m\^{e}me spectre
que celui de la th\'{e}orie de la corde h\'{e}t\'{e}rotique sur $T^{3}$.

Rappelons qu'\`{a} dix dimensions, la corde h\'{e}t\'{e}rotique a les champs
bosoniques non massifs suivant: une m\'{e}trique, une 2-forme $B$, un
dilaton $\phi $ et des champs de jauge non ab\'{e}liens de groupe de
structure $SO(32)$ ou $E_{8}\times {E_{8}}$. De plus, elle poss\`{e}de 16
supercharges globales qui restent pr\'{e}serv\'{e}es lors de la
compactification sur un tore $T^{3}$. Ainsi une m\'{e}trique plate sur $%
T^{3} $ donne lieu \`{a} six scalaires non massifs. Ceux-ci sont joints par
trois de plus provenant du champs $B$ et qui sont caract\'{e}ris\'{e}s par
les trois 2-formes harmoniques ind\'{e}pendantes dans $T^{3}$. D'autre part,
pour que les champs de jauge sur $T^{3}$ soient supersym\'{e}triques, il
faut que leurs champs de forces s'annulent. Autrement dit il faut qu'ils
aient des connections plates surtout qu'ils sont param\'{e}tris\'{e}s par
les lignes de Wilson autour des trois cercles ind\'{e}pendants du tore.
Ainsi les bosons de jauge fournissent apr\`{e}s la compactification toro%
\"{\i}dale $16\times 3$ modules associ\'{e}es aux lignes de Wilson. Dans
notre cas le tore maximal, du groupe de jauge, duquel provient les
connections plates des champs est ${U(1)}^{16}$.

Il est clair \`{a} pr\'{e}sent que le nombre total des scalaires est 58. D'o%
\`{u} localement selon Narain, l'espace des modules a la m\^{e}me forme que
celui de (\ref{e1}) avec ${\bf R}^{+}$ qui param\'{e}trise dans ce cas les
valeurs possibles de la constante de couplage de la corde. Ajoutons que le
groupe de jauge \`{a} sept dimensions (pour la m\'{e}trique et le champ $B $%
) est un sous-groupe de $SO(32)$ ou de $E_{8}\times {E_{8}}$ qui commute
avec les connections plates de $T^{3}$. Plus exactement ce groupe de jauge
est le groupe ab\'{e}lien ${U(1)}^{16}$. En plus, les trois 1-formes
harmoniques de $T^{3}$ impliquent 3 champs de jauge $U(1)$ produit par le
champ $B$ auxquels s'ajoutent trois autres donn\'{e}s par la m\'{e}trique.

Finalement, nous concluons que dans un point g\'{e}n\'{e}rique de l'espace
des modules ${\bf {\cal M}}$, les th\'{e}ories de supergravit\'{e} \`{a}
basse \'{e}n\'{e}rgie r\'{e}sultantes soit de la compactification de la th%
\'{e}orie-M sur $K3$ ou de la corde h\'{e}t\'{e}rotique sur $T^{3}$ sont les
m\^{e}mes.

\subsection{{\protect\LARGE \ }Th\'{e}orie-M \`{a} quatre dimensions}

Comme nous l'avons d\'{e}j\`{a} mentionn\'{e}, la th\'{e}orie-M est consid%
\'{e}r\'{e}e comme ''m\`{e}re'' des cinq th\'{e}ories des supercordes. Par
cons\'{e}quent, il est \'{e}vident qu'elle soit duale \`{a} la corde h\'{e}t%
\'{e}rotique \`{a} sept dimensions aussi bien qu'elle contribue dans la
description des diff\'{e}rentes th\'{e}ories de supercordes dans des
dimensions inf\'{e}rieures \`{a} dix. De ce fait, plusieurs questions
s'imposent et persistent surtout que notre objectif fondamental est
d'atteindre les quatre dimensions de l'espace-temps ${\bf R}^{1,3}$.

Ainsi dans le but de compl\'{e}ter notre \'{e}tude tout en r\'{e}pondant aux
diff\'{e}rentes questions, nous commen\c{c}ons tout d'abord par d\'{e}voiler
certaines des caract\'{e}ristiques et des propri\'{e}t\'{e}s de la vari\'{e}t%
\'{e} compacte qui nous permettra d'aboutir un mod\`{e}le \`{a} quatre
dimensions avec un nombre minimal de supercharges.

\subsubsection{{\protect\Large \ }{\bf Variations supersym\'{e}triques}}

Puisque la th\'{e}orie-M est une th\'{e}orie supersym\'{e}trique, alors il
est naturel de chercher ses vides supersym\'{e}triques. Dans le cas de la th%
\'{e}orie classique, ceci revient juste \`{a} trouver les conditions pour
lesquelles les variations supersym\'{e}triques des trois champs, qui
constituent le supermultiplet gravitationnel, s'annulent. Mais, le fait que
sous la transformation de Lorentz la valeur moyenne dans le vide du champs $%
\psi $ doit \^{e}tre nulle incite les variations des champs $g$ et $C$ \`{a}
s'annuler automatiquement. \newline
Ainsi, il reste \`{a} attribuer aux champs $C$ et \`{a} la m\'{e}trique $g$
des valeurs pour les quelles la variation de $\psi $ soit nulle: 
\begin{eqnarray}
{\delta \psi _{N}}&\equiv& {\nabla _{N}\eta }  \nonumber \\
&+&{\frac{1}{288}\left( {\Gamma _{N}^{PQRS}F_{PQRS}}-{6\Gamma ^{PQR}F_{NPQR}}%
\right) \eta }  \nonumber \\
&=&0.
\end{eqnarray}
Il est clair \`{a} pr\'{e}sent que la fa\c{c}on la plus simple pour r\'{e}%
soudre ces \'{e}quations est d'octroyer \`{a} $F$ la valeur z\'{e}ro $(F=0).$
Dans ce cas, nous devons chercher une vari\'{e}t\'{e} \`{a} onze dimensions
avec une m\'{e}trique $g$ qui admet un spineur cavariantiquement constant
(ou un spineur parall\`{e}le): 
\begin{equation}
{\nabla _{N}{\eta }}=0
\end{equation}
Cette \'{e}quation peut \^{e}tre r\'{e}\'{e}crite sous une forme plus
symbolique 
\begin{equation}
{\nabla _{g}{\eta }}=0,
\end{equation}
o\`{u} $\nabla _{g}$ est la connection de Levi-Civita construite \`{a}
partir de $g$. \newline
Ainsi, les solutions pour ces conditions peuvent \^{e}tre class\'{e}es via
le groupe d'holonomie de la connection $\nabla _{g}.$

\subsubsection{{\protect\large \ }{\bf Groupe d'holonomie}}

Le groupe d'holonomie d'une connection agissant sur un champ comme $\eta $
ou sur un champ vecteur, se comprend en terme du transport parall\`{e}le.%
\newline
Pour une vari\'{e}t\'{e} Riemanienne \`{a} onze dimensions, consid\'{e}rons
une courbe ferm\'{e}e autour de laquelle le champ est transport\'{e}. Alors
dans le cas o\`{u} la connection est une connection de Levi-Civita, le champ
revient \`{a} son point de d\'{e}part gr\^{a}ce \`{a} une rotation du groupe 
$SO(1,n-1)$ qui est exactement $SO(1,10)$ pour la th\'{e}orie-M. Le groupe g%
\'{e}n\'{e}r\'{e} par l'ensemble des rotations responsables du transport
parall\`{e}le du champs autour des courbes ferm\'{e}es de la vari\'{e}t\'{e}
est nomm\'{e} groupe d'holonomie de $\nabla _{g}$. Ce groupe est not\'{e} $%
Hol(g)$ puisqu'il d\'{e}pend du choix de la m\'{e}trique $g$. \newline
En effet, si nous n'imposons aucune condition sur g, alors pour notre mod%
\`{e}le $Hol(g)=SO(1,10)$. Par contre pour un choix particulier de g, $%
Hol(g) $ est un sous groupe propre de $SO(1,10)$.

Or, rappelons que notre but majeur dans cette section est d'atteindre un mod%
\`{e}le se propageant dans l'espace \`{a} quatre dimensions \`{a} partir de
la th\'{e}orie-M \`{a} onze dimensions. C'est pourquoi, nous allons
consacrer la suite de cette section \`{a} l'\'{e}tude de la vari\'{e}t\'{e}
qui nous permettra d'aboutir \`{a} notre fin.

\subsubsection{{\protect\Large \ }{\bf Vari\'{e}t\'{e} d'holonomie }$G_{2} $}

A ce stade, il est clair que seule la compactification sur une vari\'{e}t%
\'{e} compacte de dimension sept $X$ peut fournir un mod\`{e}le dans ''notre
espace-temps'': 
\begin{equation}
R^{1,10}\longrightarrow R^{1,3}\times X
\end{equation}
En particulier, la m\'{e}trique $g$ \`{a} onze dimensions s'\'{e}crit comme
produit de deux m\'{e}triques l'une de $X$ not\'{e}e $g^{^{\prime }}(X)$ et
l'autre de Minkowski $R^{1,3}$.\newline
Dans ce cas, le groupe de Lorentz de l'espace-temps \`{a} onze dimensions se
brise explicitement sous la forme suivante: 
\begin{equation}
SO(1,10)\longrightarrow SO(1,3)\times SO(7)
\end{equation}
o\`{u} $SO(1,3)$ est le groupe de Lorentz \`{a} quatre dimensions.\newline
Apr\`{e}s cette brisure, les conditions supersym\'{e}triques sont
satisfaites si la m\'{e}trique $g^{^{\prime }}(X)$ admet un spineur $\theta $
ob\'{e}issant la relation: 
\begin{equation}
{\nabla _{g^{^{\prime }}(X)}{\theta }}=0  \label{e2}
\end{equation}
en choisissant 
\begin{equation}
\eta =\theta \otimes \epsilon
\end{equation}
o\`{u} $\epsilon $ est une base de spineurs constants dans l'espace de
Minkowski. \newline
Dans ce cas, la condition (\ref{e2}) impose au groupe d'holonomie de la vari%
\'{e}t\'{e} compacte $X$ de dimension sept, que nous notons $Hol(g^{^{\prime
}}(X)),$ d'\^{e}tre un sous groupe de $SO(7)$. \newline
Ce fait permet \`{a} la vari\'{e}t\'{e} $X$ d'\^{e}tre r\'{e}alis\'{e}e de
plusieurs mani\`{e}res impliquant ainsi diff\'{e}rents mod\`{e}les \`{a}
quatre dimensions. A titre d'exemple, consid\`{e}rons les trois r\'{e}%
alisations suivantes:

\begin{itemize}
\item $X=T^{7}$ qui a pour groupe d'holonomie $U(1)^{7}$. Le mod\`{e}le
obtenu \`{a} quatre dimensions poss\'{e}de 32 supercharges d'o\`{u} ${\cal N}%
=8.$

\item $X=K3\times T^{3}$ le mod\`{e}le r\'{e}sultant dans ce cas a 16
supercharges c\`{a}d ${\cal N}=4.$

\item $X=CY^{3}\times S^{1}$ donne \`{a} quatre dimensions une th\'{e}orie
avec 8 supercharges ${\cal N}=2.$
\end{itemize}

Mais notre objectif est de retrouver un mod\`{e}le \`{a} quatre dimensions
avec un nombre de supercharges minimales (c\`{a}d 4 charges supersym\'{e}%
triques ce qui implique ${\cal N}=1$), apr\`{e}s la compactification de la th%
\'{e}orie-M \`{a} onze dimensions poss\'{e}dant 32 supercharges. Alors le
choix de la vari\'{e}t\'{e} $X$ doit \^{e}tre tr\`{e}s particulier.\newline
Parmi toutes les r\'{e}alisations possibles de $X$, seule une vari\'{e}t\'{e}
ayant le groupe d'holonomie ${\bf G_{2}}$ qui est le sous groupe propre
maximal de $SO(7)$ peut avoir une repr\'{e}sentation spinorielle contenant
un singlet \cite{ref7}. \newline
De fa\c{c}on explicite 
\begin{eqnarray}
SO(1,10) &\longrightarrow &SO(1,3)\times SO(7)  \nonumber \\
32 &\longrightarrow &(4,8)  \label{e3}
\end{eqnarray}
avec 8 est la repr\'{e}sentation spinorielle de $SO(7)$ qui se d\'{e}compose
en un singlet et une repr\'{e}sentation fondamentale de ${\bf G_{2}}$ que
nous \'{e}crivons: 
\begin{equation}
8\longrightarrow 1+7.
\end{equation}
Ainsi (\ref{e3}) se r\'{e}ecrit comme suit: 
\begin{eqnarray}
SO(1,10) &\longrightarrow &SO(1,3)\times SO(7)  \nonumber \\
32 &\longrightarrow &(4,1)\oplus (4,7)
\end{eqnarray}
Nous d\'{e}duisons alors que seule la compactification de la th\'{e}orie-M
sur une vari\'{e}t\'{e} compacte de dimension sept ayant un groupe
d'holonomie ${\bf G_{2}}$ permet de r\'{e}aliser un mod\`{e}le \`{a} quatre
dimensions poss\'{e}dant un nombre minimal de supercharges, plus exactement
quatre supercharges d'o\`{u} ${\cal N}=1$ et pas plus.

\subsubsection{{\protect\Large \ }Propri\'{e}t\'{e}s des vari\'{e}t\'{e}s
d'holonomie $G_{2}$}

Il est vrai que la m\'{e}trique $g^{^{\prime }}(X)$ choisit pr\'{e}c\'{e}%
demment admet $\theta $ comme spineur parall\`{e}le. Mais ceci n'emp\^{e}che
de construire d'autres champs de $X$ qui soient covariantiquement constants 
\cite{ref7}.\newline
Ainsi toutes $p$-formes ayant les composantes suivantes: 
\begin{equation}
\theta ^{T}\Gamma _{i_{1}...i_{p}}\theta
\end{equation}
peuvent \^{e}tre parall\`{e}les par respect de $\nabla _{g^{^{\prime }}(X)}$%
. Cependant ces $p$-formes ne sont non nulles que pour des valeurs bien
distinctives de $p$ et qui ne sont autres que: $0,3,4$ et $7.$ Ceci est d%
\^{u} au fait que la d\'{e}composition de la repr\'{e}sentation antisym\'{e}%
trique de $SO(7)$ en terme des repr\'{e}sentations de ${\bf G}_{2}$
contenant le singlet n'est valable que pour ces quatre valeurs de $p$.
D'autre part, en plus de la 0-forme qui n'est autre qu'une constante dans $X$
et la 7-forme qui est sa forme volume, la 3-forme ainsi que son duale la
4-forme jouent un r\^{o}le primordial. Selon \cite{ref9,ref10}, la 3-forme
est consid\'{e}r\'{e}e comme l'ensemble des structures constantes pour l'alg%
\`{e}bre d'octonion, permettant de ce fait \`{a} ${\bf G}_{2}$ d'\^{e}tre le
groupe d'automorphisme de cette alg\`{e}bre.\newline

Pour mieux explicite cette id\'{e}e, nous allons \'{e}tudier de pr\`{e}s
l'importance de la 3-forme en consid\'{e}rant tout d'abord, un syst\`{e}me
de coordonn\'{e}es $(x_{1},...,x_{7})$ dans un espace plat ${\bf R}^{7}.$
Notons la forme ext\'{e}rieur $dx_{i}\wedge ...\wedge dx_{l}$ de ${\bf R}%
^{7} $ par $dx_{i...l}$ et d\'{e}finissons une m\'{e}trique $%
g^{^{\prime}}_{0} $, une 3-forme $\varphi_{0}$ et son dual de Hodge la
quatre forme $\star\varphi_{0}$ comme suit: 
\begin{eqnarray}
g^{^{\prime}}_{0}&=&dx_{1}^{2}+....+dx_{7}^{2}  \nonumber \\
\varphi_{0}&=&dx_{123}+dx_{145}+dx_{167}+dx_{246}  \nonumber \\
&-&dx_{257}-dx_{356}-dx_{347}  \nonumber \\
\star\varphi_{0}&=&dx_{4567}+dx_{2367}+dx_{2345}+dx_{1357}  \nonumber \\
&-&dx_{1346}- dx_{1256}-dx_{1247}
\end{eqnarray}
Alors, le sous-groupe de $Gl(7,{\bf R})$ qui pr\'{e}serve $\varphi_{0}$
n'est autre que le groupe de Lie exceptionnel ${\bf G_{2}.}$ Il pr\'{e}serve
aussi la 4-forme $\star\varphi_{0}$ ainsi que $g^{^{\prime}}_{0} $ la m\'{e}%
trique euclidienne. \newline
Le groupe de Lie ${\bf G_{2}}$ est compact, semi-simple et de dimension 14.
De plus, c'est le sous-groupe maximal de $SO(7)$. Dans ce cas, $\varphi_{0}$
et $\star\varphi_{0}$ d\'{e}finissent la structure-${\bf G_{2}}$ dans ${\bf R%
}^{7}.$ \newline
Par la suite, int\'{e}ressons-nous au cas o\`{u} la vari\'{e}t\'{e} $X$ de
dimension sept est courb\'{e}e. Alors la g\'{e}om\'{e}trie est d\'{e}termin%
\'{e}e par une 3-forme $\varphi$ stable par ${\bf G_{2},}$ de sorte que tout
espace tangent $T_{p}(X)$ \`{a} $X $ en un point $p$ admet un isomorphisme
avec ${\bf R}^{7}.$ Celui-ci identifie $\varphi$ et la m\'{e}trique $%
g^{^{\prime}}(X)$ avec $\varphi_{0}$ et $g_{0}$ respectivement. D'autre
part, pour l'alg\`{e}bre d'Octonion o\`{u} 
\begin{equation}
{{{\bf \widetilde{o}}}}=x^{0}1 + x^{a}i_{a},
\end{equation}
le groupe exceptionnel ${\bf G_{2}}$ est son groupe d'automorphisme.
Effectivement pour une 3-forme $\varphi$ qui est d\'{e}finit localement, les 
\'{e}l\'{e}ments $i_{a}$ satisfont la formule 
\begin{equation}
i_{a}i_{b}=-\delta_{ab}+\varphi_{abc} i_{c}
\end{equation}
suite \`{a} laquelle $ImO$ qui est la partie imaginaire de l'alg\`{e}bre
d'Octonion est consid\'{e}r\'{e}e comme une copie de l'espace tangent en un
point de la vari\'{e}t\'{e} $X$.\newline
Par ailleurs, vu l'importance de cette 3-forme $\varphi,$ la structure-${\bf %
G_{2}}$ est not\'{e}e par la paire $(\varphi,g^{^{\prime}}).$ Par cons\'{e}%
quent, la vari\'{e}t\'{e} $X$ d'holonomie ${\bf G_{2}}$ est d\'{e}finit par
le triplet $(X,\varphi,g^{^{\prime}})$ o\`{u} $X$ est la vari\'{e}t\'{e} de
dimension sept dans laquelle la structure-${\bf G_{2}}$ est de torsion libre%
\footnote{%
Dire que la torsion $\nabla\varphi$ est libre revient \`{a} ce qu'elle v\'{e}%
rifie la condition $\nabla\varphi=0$ sur $X$ avec $\nabla$ est la connection
de Levi-Civita associ\'{e}e \`{a} la m\'{e}trique $g^{^{\prime}}(X)$.}.%
\newline
Dans ce cas, on a les propri\'{e}t\'{e}s suivantes:

\begin{enumerate}
\item {\ Le triplet $(X,\varphi ,g^{^{\prime }})$ est une vari\'{e}t\'{e}
d'holonomie ${\bf G_{2}}$ si $X$ est une vari\'{e}t\'{e} de dimension sept
et $(\varphi ,g^{^{\prime }})$ est la structure-${\bf G_{2}}$ de torsion
libre dans $X. $}

\item Si $g^{^{\prime }}$ a une holonomie $Hol(g^{^{\prime }}(X))=G_{2}$
alors $g^{^{\prime }}$ est Ricci-plate.

\item {\ La vari\'{e}t\'{e} $X$ de dimension sept et d'holonomie ${\bf G_{2}}
$ poss\'{e}de un groupe fondamental $\pi _{1}$ qui est fini. Par cons\'{e}%
quent son premier nombre de Betti $b_{1}(X)=0$. Cependant le troisi\`{e}me
nombre de Betti $b_{3}(X)$ repr\'{e}sente la dimension de l'espace des
modules de $g^{^{\prime }}(X)$.}

\item La compactification sur une vari\'{e}t\'{e} d'holonomie ${\bf G_{2}}$
pr\'{e}serve $\frac{1}{8}$ des charges supersym\'{e}triques initiales.
\end{enumerate}

\subsubsection{{\protect\Large \ }{\bf R}\'{e}duction de Kaluza-Klein}

Nous avons vu jusqu'\`{a} pr\'{e}sent que dans le cadre de la
compactification de la th\'{e}orie-M vers $D=4$, seule une vari\'{e}t\'{e} $%
X $ r\'{e}guli\`{e}re d'holonomie ${\bf G_{2}}$ assure l'obtention d'un mod%
\`{e}le \`{a} quatre dimensions avec ${\cal N}=1$. Nous compl\'{e}tons cette
section par une description explicite de ce mod\'{e}le en utilisant
l'analyse de Kaluza-Klein \cite{ref11,ref12,ref13} pour le champs antisym%
\'{e}trique $C$ et la m\'{e}trique $g^{^{\prime }}$ .

{{\bf a) L'analyse de Kaluza-Klein pour le champs }$C$}

Choisissons deux bases des formes harmoniques de $H^{2}(X)$ et de $H^{3}(X)$%
: 
\begin{equation}
\{\beta ^{\gamma };\gamma =1,...,b_{2}(X)\}
\end{equation}
{\text{e}t} 
\begin{equation}
\{\omega ^{a};a=1,...,b_{3}(X)\}.
\end{equation}
Du fait que $b_{1}(X)=0$, alors il n'y aura pas de sommation sur les
1-formes harmoniques de $X$. Ainsi l'ansatz pour $C$ ne produit pas des
2-formes non massives \`{a} $D=4$. Par cons\'{e}quent, l'ansatz de
Kaluza-Klein pour le champs $C$ s'\'{e}crit dans les deux bases $\omega $ et 
$\beta $ comme suit: 
\begin{equation}
C={\sum_{a}{\omega ^{a}\phi _{a}(y)}}+{\sum_{\gamma }{\beta ^{\gamma
}A_{\gamma }(y)}}
\end{equation}
Les $A$ repr\'{e}sentent les 1-formes de l'espace de Minkowski impliquant $%
b_{2}(X)$ champs de jauge ab\'{e}liens, alors que les $\phi _{a}$ sont les
champs scalaires d\'{e}pendant des coordonn\'{e}es $y$ de l'espace de
Minkowski \`{a} D=4.

Bien que le champs $C$ produise $b_{3}(X)$ scalaires r\'{e}els non massifs,
il faut s'attendre selon la supersym\'{e}trie ${\cal N}=1$ \`{a} $D=4$ \`{a}
ce que l'analyse de Kaluza-Klein pour $g^{^{\prime}}(X)$ donne $b_{3}(X)$
scalaires suppl\'{e}mentaires \`{a} quatre dimensions. D'ailleurs pour l'alg%
\`{e}bre supersym\'{e}trique ${\cal N}=1,$ toutes ses repr\'{e}sentations
qui contiennent un scalaire r\'{e}el non massif, suscitent \`{a} avoir au
total deux scalaires se combinant en scalaire complexe.

{{\bf b) L'analyse de Kaluza-Klein pour }$g^{^{\prime }}(X)$}

La m\'{e}trique $g^{^{\prime }}(X)$ de la vari\'{e}t\'{e} $X$ d'holonomie $%
{\bf G_{2}}$ ob\'{e}it aux \'{e}quations du vide d'Einstein: 
\begin{equation}
R_{ij}[g^{^{\prime }}(X)]=0  \label{l2}
\end{equation}
avec $R_{ij}$ est le tenseur de Ricci. \newline
Alors l'obtention du spectre des modes z\'{e}ro provenant de la m\'{e}trique 
$g^{^{\prime }}(X)$ exige de lui chercher des fluctuations qui satisfont
aussi aux \'{e}quations (\ref{l2}).\newline
Attribuons \`{a} ces fluctuations de la m\'{e}trique la forme 
\begin{equation}
g_{ij}^{^{\prime }}(x)+\delta g_{ij}^{^{\prime }}(x,y)
\end{equation}
o\`{u} $\delta g_{ij}^{^{\prime }}$ d\'{e}pendent des coordonn\'{e}es $y$ de
l'espace de Minkowski \`{a} $D=4$ ainsi que des coordonn\'{e}es $x$ de la
vari\'{e}t\'{e} d'holonomie ${\bf G_{2}}$. Ensuite, utilisons l'ansatz de
Kaluza-Klein pour les fluctuations donn\'{e} par la forme 
\begin{equation}
\delta g_{ij}^{^{\prime }}=h_{ij}(x)\rho (y).
\end{equation}
Ces fluctuations produisent des champs scalaires non massifs \`{a} quatre
dimensions.\newline
D'autre part, pour une vari\'{e}t\'{e} de dimension sept et d'holonomie $%
SO(7)$, les $h_{ij}$ sont des tenseurs sym\'{e}triques d'ordre 2. Ils se
transforment via la repr\'{e}sentation de dimension {\bf 27}. Cette derni%
\`{e}re reste irr\'{e}ductible m\^{e}me dans le cas de ${\bf G_{2}}$. Par
ailleurs dans une vari\'{e}t\'{e} d'holonomie ${\bf G_{2}}$, les 3-formes
qui sont dans la repr\'{e}sentation {\bf 35} de $SO(7)$ se d\'{e}composent
sous l'action de ${\bf G_{2}}$ comme suit: 
\begin{equation}
{\bf 35}\longrightarrow {\bf 1+7+27}.
\end{equation}
Permettant ainsi aux $h_{ij}$ d'\^{e}tre consid\'{e}r\'{e}s comme des
3-formes de $X$. Or, puisque $\varphi $ est une 3-forme dans la repr\'{e}%
sentation triviale, alors les $\omega $ sont des 3-formes dans la m\^{e}me
repr\'{e}sentation que $h_{ij}$. D'o\`{u} les fluctuations de la structure-$%
{\bf G_{2}}$ s'\'{e}crivent: 
\begin{eqnarray}
\varphi ^{^{\prime }} &=&\varphi +\delta \varphi  \nonumber \\
&=&\varphi +{\sum_{a}{\omega ^{a}(x)\rho _{a}(y)}}.
\end{eqnarray}
Par cons\'{e}quent, les champs scalaires non massifs provenant des
fluctuations de la m\'{e}trique de $X$ sont exactement au nombre de $%
b_{3}(X) $. Ainsi la combinaison des $\phi $ et $\rho $ donne $b_{3}(X)$
scalaires non massifs not\'{e}s $\Phi ^{a}(y)$. De plus, les
superpartenaires fermioniques pour tout ces champs r\'{e}sultants sont
fournit par le biais de la r\'{e}duction de Kaluza-Klein pour le gravitino.%
\newline

En conclusion, la compactification de la th\'{e}orie-M sur une vari\'{e}t%
\'{e} $X$ r\'{e}guli\`{e}re d'holonomie ${\bf G_{2}}$ est une fa\c{c}on
naturelle pour obtenir un mod\`{e}le \`{a} $D=4$ ${\cal N}=1$ ayant un
groupe de jauge ab\'{e}lien. Ce mod\`{e}le est d\'{e}crit par une th\'{e}%
orie de supergravit\'{e} ${\cal N}=1$ coupl\'{e}e \`{a} $b_{2}(X)$
multiplets vectoriels ab\'{e}liens et \`{a} $b_{3}(X)$ multiplets chiraux
non massifs. Pourtant cette compactification peut bien offrir une th\'{e}%
orie plus int\'{e}ressante poss\'{e}dant un groupe de jauge non ab\'{e}lien
et de la mati\`{e}re chirale, mais \`{a} condition que la vari\'{e}t\'{e} $X$
admette un certain type de singularit\'{e}s. Motiv\'{e}s par cette id\'{e}e 
\cite{refA}, la construction de cette vari\'{e}t\'{e} ainsi que l'\'{e}tude
de la physique localis\'{e}e au voisinage des singularit\'{e}s de l'espace
d'holonomie ${\bf G_{2}}$ fera l'objet de la section suivante.

\section{Th\'{e}orie-M au Voisinage des Singularit\'{e}s}

Le but principal de cette section est de proc\'{e}der \`{a} la construction
d'une vari\'{e}t\'{e} d'holonomie ${\bf G}_{2}$ qui nous attribue un mod\`{e}%
le \`{a} quatre dimensions, poss\'{e}dant un groupe de jauge non ab\'{e}lien
ainsi que de la mati\`{e}re chirale. En premier temps, nous utilisons comme
argument de d\'{e}part la dualit\'{e} entre la th\'{e}orie-M compactifi\'{e}%
e sur $K3$ et la corde h\'{e}t\'{e}rotique sur $T^{3}.$ Cette \'{e}tude de
dualit\'{e} nous pr\'{e}sente une mani\`{e}re g\'{e}om\'{e}trique pour
extraire les sym\'{e}tries de jauge non ab\'{e}liennes dans le mod\`{e}le de
la th\'{e}orie-M \`{a} partir de l'aspect singulier de $K3.$ Ensuite nous r%
\'{e}alisons en d\'{e}tail la vari\'{e}t\'{e} $X$ dont les singularit\'{e}s
vont nous permettre d'obtenir un mod\`{e}le de physique de particules \`{a}
quatre dimensions.

\subsection{Vari\'{e}t\'{e} $K3$\ et les singularit\'{e}s $ADE$}

L'une des cons\'{e}quences de l'\'{e}tude compl\`{e}te de la dualit\'{e}
entre la th\'{e}orie-M et la corde h\'{e}t\'{e}rotique est que de m\^{e}me
que pour cette derni\`{e}re, la th\'{e}orie-M doit aussi poss\'{e}der une sym%
\'{e}trie non ab\'{e}lienne \cite{ref5}. Cette sym\'{e}trie est garantit par
des points sp\'{e}ciaux de l'espace des modules de $K3$. Plus pr\'{e}cis\'{e}%
ment, ceux o\`{u} la vari\'{e}t\'{e} $K3$ d\'{e}veloppe des singularit\'{e}s
orbifold. De ce fait, nous concentrons notre int\'{e}r\^{e}t dans cette
sous-section \`{a} une r\'{e}alisation g\'{e}om\'{e}trique bien particuli%
\`{e}re de $K3$. Exactement, celle o\`{u} cette vari\'{e}t\'{e} riemannienne
de dimension quatre est d\'{e}crite localement par l'orbifold ${{\bf R}^{4}}/%
{\Gamma }$ o\`{u} $\Gamma $ est un sous-groupe discret et fini de $SO(4)$.
Mais sachant que la supersym\'{e}trie est compl\`{e}tement non bris\'{e}e
sur tout l'espace des modules dans le cas de la corde h\'{e}t\'{e}rotique,
alors les singularit\'{e}s orbifold de $K3$ doivent \^{e}tre choisies de
sorte que la supersym\'{e}trie soit pr\'{e}serv\'{e}e l\`{a} aussi. Ceci se
traduit par la condition que $\Gamma $ est un sous-groupe fini de $%
SU(2)\subset SO(4).$ \newline
Par passage aux coordonn\'{e}es complexes ${\bf C^{2}}\equiv {\bf R^{4}}$,
tout point de ${\bf C^{2}}$ est repr\'{e}sent\'{e} par un vecteur \`{a} deux
composantes sur lequel $SU(2)$ agit de fa\c{c}on standard comme suit: 
\begin{equation}
\left( 
\begin{array}{c}
u \\ 
v%
\end{array}
\right) \qquad \rightarrow \qquad \left( 
\begin{array}{cc}
a & b \\ 
c & d%
\end{array}
\right) \left( 
\begin{array}{c}
u \\ 
v%
\end{array}
\right)
\end{equation}
o\`{u} les sous-groupes finis de $SU(2)$ ont une classification d\'{e}crite
en termes des alg\`{e}bres de Lie semi-simples et simplement lac\'{e}es \cite%
{ref7}: ${\rm A_{n}}$, ${\rm D_{k}}$, ${\rm E_{6}}$, ${\rm E_{7}}$ et ${\rm %
E_{8}}$ .\newline
Dans ce contexte, les trois sous groupes exceptionnels correspondent aux
trois groupes de Lie exceptionnel de type-${\rm E}.$ Cependant ${\rm A_{n}}$
est li\'{e}e \`{a} $SU(n+1)$ alors que ${\rm D_{k}}$ est associ\'{e}e \`{a} $%
SO(2k).$\newline
D'autre part, ces sous-groupes que nous notons par $\Gamma _{{\rm A_{n}}}$, $%
\Gamma _{{\rm D_{k}}}$ et $\Gamma _{{\rm E_{i}}}$ sont d\'{e}crits
explicitement comme suit:\newline

\begin{itemize}
\item $\Gamma _{{\rm A_{n-1}}}$ est isomorphe \`{a} ${\bf Z_{n}}$ (le groupe
cyclique d'ordre $n$) et il est g\'{e}n\'{e}r\'{e} par: 
\begin{equation}
\left( 
\begin{array}{cc}
e^{\frac{2i\pi }{n}} & 0 \\ 
0 & e^{\frac{-2i\pi }{n}}%
\end{array}
\right)
\end{equation}

\item $\Gamma _{{\rm D_{k}}}$ est isomorphe \`{a} ${\cal D}_{k-2}$ (le
groupe dih\'{e}dral binaire d'ordre $4k-8$) et admet 2 g\'{e}n\'{e}rateurs $%
\alpha $ et $\sigma $ donn\'{e}s par: 
\begin{equation}
\left( 
\begin{array}{cc}
e^{\frac{i\pi }{k-2}} & 0 \\ 
0 & e^{\frac{-i\pi }{k-2}}%
\end{array}
\right) ;\qquad \left( 
\begin{array}{cc}
0 & i \\ 
i & i%
\end{array}
\right)
\end{equation}

\item $\Gamma _{{\rm E_{6}}}$ est isomorphe \`{a} ${\cal T}$ (le groupe t%
\'{e}trahedral binaire d'ordre $24$) et qui poss\`{e}de deux g\'{e}n\'{e}%
rateurs de la forme: 
\begin{equation}
\left( 
\begin{array}{cc}
e^{\frac{i\pi }{2}} & 0 \\ 
0 & e^{\frac{-i\pi }{2}}%
\end{array}
\right) ;\qquad \left( 
\begin{array}{cc}
e^{\frac{2i\pi 7}{8}} & e^{\frac{2i\pi 7}{8}} \\ 
e^{\frac{2i\pi 5}{8}} & e^{\frac{2i\pi }{8}}%
\end{array}
\right)
\end{equation}

\item $\Gamma _{{\rm E_{7}}}$ est isomorphe \`{a} ${\rm O}$ (le groupe octoh%
\'{e}dral binaire d'ordre 48) ayant pour g\'{e}n\'{e}rateurs les deux de $%
\Gamma _{{\rm E_{6}}}$ et un troisi\`{e}me qui est:

\begin{equation}
\left( 
\begin{array}{cc}
e^{\frac{2i\pi }{8}} & 0 \\ 
0 & e^{\frac{2i\pi 7}{8}}%
\end{array}
\right)
\end{equation}

\item Finalement, $\Gamma _{{\rm E_{8}}}$ est isomorphe \`{a} ${\rm I} $ (le
groupe icosah\'{e}dral d'ordre 120) avec deux g\'{e}n\'{e}rateurs: 
\begin{eqnarray}
&&\left( 
\begin{array}{cc}
-e^{\frac{2i\pi 3}{5}} & 0 \\ 
0 & -e^{\frac{2i\pi 7}{8}}%
\end{array}
\right) ,  \nonumber \\
&&\left( 
\begin{array}{cc}
\frac{{{e^{\frac{2i\pi }{5}}}+{e^{\frac{-2i\pi }{5}}}}}{{e^{\frac{2i\pi 2}{5}%
}}-{e^{\frac{2i\pi 3}{5}}}} & \frac{ 1}{{e^{\frac{2i\pi 2}{5}}}-{e^{\frac{%
2i\pi 3}{5}}}} \\ 
\frac{ 1}{{e^{\frac{2i\pi 2}{5}}}-{e^{\frac{2i\pi 3}{5}}}} & \frac{{{-e^{%
\frac{2i\pi }{5}}}-{e^{\frac{-2i\pi }{5}}}}}{{e^{\frac{2i\pi 2}{5}}}-{e^{%
\frac{2i\pi 3}{5}}}}%
\end{array}
\right)
\end{eqnarray}
\end{itemize}

Puisque seule la physique au voisinage des singularit\'{e}s orbifold de $K3$
nous int\'{e}resse, alors nous allons nous restreindre dans ce qui suit \`{a}
une \'{e}tude locale de la th\'{e}orie-M sur ${{\bf C}^{2}}/{\Gamma }\times 
{\bf R}^{1,6}$. De plus, nous allons choisir le sous groupe $\Gamma _{{\rm %
A_{1}}}$ parmi tout les autres pour des raisons de simplicit\'{e}. Ainsi
dans ce cas, nous avons $\Gamma _{{\rm A_{1}}}$ qui est isomorphe \`{a} $%
{\bf Z_{2}},$ est en fait le centre de $SU(2)$. Son g\'{e}n\'{e}rateur agit
sur ${\bf C}^{2}$ comme suit: 
\begin{equation}
\left( 
\begin{array}{c}
u \\ 
v%
\end{array}
\right) \qquad \rightarrow \qquad \left( 
\begin{array}{c}
-u \\ 
-v%
\end{array}
\right)
\end{equation}
Alg\'{e}briquement, param\'{e}trisons ${{\bf C}^{2}}/{\Gamma _{{\rm A_{1}}}}$
en termes des coordonn\'{e}es de ${{\bf C}^{2}}$ invariants par ${\Gamma _{%
{\rm A_{1}}}}$ que nous notons: $u^{2},v^{2}$ et $uv$. Ensuite, nous donnons
une transformation: 
\begin{equation}
{{\bf C}^{2}}/{\Gamma _{{\rm A_{1}}}}\longrightarrow {{\bf C}^{3}}
\end{equation}
et nous d\'{e}finissons les nouvelles coordonn\'{e}es complexes: 
\begin{eqnarray}
x &=&u^{2}-v^{2}  \nonumber \\
y &=&iu^{2}+iv^{2}  \nonumber \\
z &=&2uv
\end{eqnarray}
qui sont invariantes par la sym\'{e}trie ${\bf Z_{2}}.$ Cette nouvelle param%
\'{e}trisation d\'{e}crit un mod\`{e}le local singulier. Ce dernier est
connu sous le nom de la singularit\'{e} $A_{1}$ dans la classification des
singularit\'{e}s des surfaces complexes: 
\begin{equation}
A_{1}:{\cal P}(x,y,z)=xy-z^{2}=0.
\end{equation}
Pour un choix ad\'{e}quat des variables $x,y,z$, l'\'{e}quation complexe
prend la forme 
\begin{equation}
x^{2}+y^{2}+z^{2}=0.  \label{q1}
\end{equation}
Dans ce cas, (\ref{q1}) d\'{e}finit l'orbifold ${{\bf C}^{2}}/{\Gamma _{{\rm %
A_{1}}}}$ comme une hypersurface singuli\`{e}re dans ${{\bf C}^{3}}$. Par
ailleurs, cette singularit\'{e} peut \^{e}tre r\'{e}solue par deux fa\c{c}%
ons:

\begin{enumerate}
\item Par la d\'{e}formation de la structure complexe en ajoutant un terme
suppl\'{e}mentaire dans l'equation alg\'{e}brique (\ref{q1}).

\item Par la d\'{e}formation de la structure de Kahler en rempla\c{c}ant le
point singulier par une sph\`{e}re.
\end{enumerate}

Or la propri\'{e}t\'{e} d'auto-miroir de la vari\'{e}t\'{e} $K3$ permet aux
deux approches des d\'{e}formations \`{a} \^{e}tre \'{e}quivalentes. Ainsi,
le traitement de la singularit\'{e} par l'une ou l'autre d\'{e}formation
conduit \`{a} des vari\'{e}t\'{e}s topologiquement identiques. Par cons\'{e}%
quent, l'\'{e}quation (\ref{q1}) apr\`{e}s la r\'{e}solution complexe
devient: 
\begin{equation}
x^{2}+y^{2}+z^{2}=\mu
\end{equation}
Il est clair que la partie r\'{e}elle de cette \'{e}quation n'est autre
qu'une 2-sph\`{e}re dont l'aire finie est d\'{e}termin\'{e}e par la partie r%
\'{e}elle de $\mu $ qui donne le rayon r\'{e}el $r.$ Par suite, lorsque r
tend vers z\'{e}ro, la sph\`{e}re se contracte \`{a} une aire nulle. D'autre
part, il se trouve que l'espace total de la d\'{e}formation de la vari\'{e}t%
\'{e} ${{\bf C}^{2}}/{\Gamma _{{\rm A_{1}}}}$ n'est autre que le fibr\'{e}
cotangent de la 2-sph\`{e}re et que nous notons par ${\rm T}^{\ast }S^{2}.$
Afin de mieux illustrer cette id\'{e}e, consid\'{e}rons les parties r\'{e}%
elles $l_{i}$ de $x,y,z$ ainsi que leurs parties imaginaires $p_{i}.$ Pour $%
\mu $ r\'{e}el, nous obtenons les deux \'{e}quations suivantes: 
\begin{eqnarray}
\sum {l_{i}^{2}}-\sum {p_{i}^{2}} &=&\mu  \nonumber \\
\sum {l_{i}p_{i}} &=&0
\end{eqnarray}
les $l_{i}$ d\'{e}crivent la sph\`{e}re $S^{2}$, alors que les $p_{i}$ param%
\'{e}trisent les directions tangentielles.

\subsection{Th\'{e}orie-M au voisinage des singularit\'{e}s $ADE$\ }

Comme nous venons de le constater, la d\'{e}formation de la singularit\'{e}
orbifold donne lieu \`{a} une 2-sph\`{e}re de rayon $r.$ Alors \`{a} sept
dimensions \cite{ref14}, un champ de jauge $U(1)$ est produit par le biais
de la r\'{e}duction de Kaluza-Klein pour le champ antisym\'{e}trique $C.$ De
plus, la m\'{e}trique d'holonomie ${\bf G_{2}}$ poss\`{e}de trois param\`{e}%
tres qui contr\^{o}lent l'aire de $S^{2}.$ Or puisque le multiplet vectoriel 
\`{a} sept dimensions contient exactement un champ de jauge et trois champs
scalaires. Nous d\'{e}duisons par analogie que le spectre non massif lorsque 
${\rm T}^{\ast }S^{2}$ est r\'{e}gulier n'est autre qu'un multiplet
vectoriel ab\'{e}lien.

D'autre part d\`{e}s que les scalaires varient vers z\'{e}ro, ce qui
signifie que la sph\`{e}re se r\'{e}tr\'{e}cit \`{a} une aire nulle, une sym%
\'{e}trie de jauge non ab\'{e}lienne appara\^{\i}t. Cette derni\`{e}re doit
son existence \`{a} la dualit\'{e} avec la corde h\'{e}t\'{e}rotique, tout
en exigeant aux bosons $W^{\pm}$ d'\^{e}tre non massifs au voisinage de la
singularit\'{e} $A_1$. Ces bosons sont \'{e}lectriquement charg\'{e}s sous
le champ de jauge $U(1)$ qui provient de la 3-forme $C.$ Or \`{a} $D=11, $
c'est la M2-brane qui est charg\'{e} sous $C.$ Portant une tension, la
dynamique de cette brane l'incite \`{a} enrouler la sph\`{e}re dans
l'espace. Ainsi \`{a} sept dimensions cet enroulement g\'{e}n\`{e}re une
particule charg\'{e}e sous $U(1)$ qui porte une masse proportionnelle \`{a}
l'aire du volume de $S^{2}.$ Plus encore, une charge $U(1)$ oppos\'{e}e \`{a}
la pr\'{e}c\'{e}dente est produite du fait que la M2-brane enroule ce
2-cycle m\^{e}me dans l'orientation oppos\'{e}e. Par suite, dans la limite
singuli\`{e}re de $K3$ o\`{u} le volume de la sph\`{e}re se contracte \`{a} z%
\'{e}ro, les deux multiplets BPS charg\'{e}s de fa\c{c}on oppos\'{e}e
deviennent non massifs. Ceux-ci provoquent l'apparition d'une sym\'{e}trie
de jauge $A_{1}=SU(2)$ \`{a} partir de $U(1)$.\newline
Par cons\'{e}quent le probl\`{e}me de la sym\'{e}trie de jauge non ab\'{e}%
lienne dans la th\'{e}orie-M est rattach\'{e}e \`{a} l'\'{e}tude de la
singularit\'{e} $A_{1}$ de $K3.$\newline

En outre, \`{a} sept dimensions la th\'{e}orie de super Yang-Mills d\'{e}%
pend seulement de son groupe de jauge. Par ailleurs en absence de la gravit%
\'{e}, la physique \`{a} basse \'{e}nergie de la th\'{e}orie-M sur ${{{\bf C}%
^{2}}/{\Gamma_{A_{1}}}}\times {\bf R}^{1,6}$ est d\'{e}crite par la th\'{e}%
orie de super Yang-Mills sur ${\cal O}\times {\bf R}^{1,6}$ avec un groupe
de jauge $SU(2)$.

Dans le but de rendre l'\'{e}tude plus compl\`{e}te, il s'est av\'{e}r\'{e} n%
\'{e}cessaire de proc\'{e}der \`{a} une g\'{e}n\'{e}ralisation. En effet en
absence de la gravit\'{e}, la physique \`{a} basse \'{e}nergie de la th\'{e}%
orie-M sur ${{{\bf C}^{2}}/{\Gamma_{ADE}}}\times {\bf R}^{1,6}$ est d\'{e}%
crite par la th\'{e}orie de super Yang-Mills sur ${\cal O}\times {\bf R}%
^{1,6}$ ayant un groupe de jauge associ\'{e} \`{a} l'une des alg\`{e}bres de
Lie $ADE$. Particuli\`{e}rement la d\'{e}formation de la singularit\'{e}
orbifold dans ${{\bf C}^{2}}/{\Gamma_{ADE}}$ contient $k$ 2-sph\`{e}res avec 
$k= rang(ADE).$ L'intersection de ces sph\`{e}res s'effectue en concorde
avec la matrice de Cartan des alg\`{e}bres de Lie $ADE.$ Par contre aux
points r\'{e}guliers de l'espace des modules, le groupe de jauge n'est autre
que $U(1)^{k}.$ \newline
En conclusion, \`{a} sept dimensions la th\'{e}orie \`{a} basse \'{e}nergie
acquiert un groupe de jauge non ab\'{e}lien li\'{e} aux $ADE,$ \`{a}
l'origine de l'espace des modules, et ceci gr\^{a}ce au m\'{e}canisme de
Higgs.

\subsection{Th\'{e}orie de jauge \`{a} quatres dimensions}

Rappelons que l'id\'{e}e principale de cette section est de retrouver un mod%
\`{e}le physique \`{a} quarte dimensions, r\'{e}sultant de la
compactification de la th\'{e}orie-M sur la vari\'{e}t\'{e} $X$ qui poss\`{e}%
de un certain type de singularit\'{e}s. Ainsi en se basant sur ce qui pr\'{e}%
c\`{e}de, nous consid\'{e}rons un espace-temps $Y^{1,6}$ \`{a} sept
dimensions le long duquel sont implant\'{e}es des singularit\'{e}s $ADE.$
Alors comme nous venons de le constater auparavant dans le contexte de la
dualit\'{e} \`{a} $D=7$, la description de la physique de la th\'{e}orie-M
au voisinage de $Y$ s'effectue en terme de la th\'{e}orie de super
Yang-Mills avec un groupe de jauge d\'{e}termin\'{e} par le type de
singularit\'{e} donn\'{e}. Par cons\'{e}quent il est clair \`{a} pr\'{e}sent
que la r\'{e}alisation de la physique de la th\'{e}orie-M compactifi\'{e}e
vers quatre dimensions, revient \`{a} l'\'{e}tude de la dynamique de la th%
\'{e}orie de jauge dans l'espace ${W}\times {{\bf R}^{1,3}}$ en absence de
la gravit\'{e} \cite{ref7,ref16,refA}. Au voisinage de cet espace qui n'est
autre que $Y,$ la vari\'{e}t\'{e} $X$ d'holonomie ${\bf G}_{2}$ admet la r%
\'{e}alisation: 
\begin{equation}
X={{{\bf C}^{2}}/{\Gamma _{ADE}}}\times {W}
\end{equation}
o\`{u} $W$ caract\'{e}rise une vari\'{e}t\'{e} de dimension trois r\'{e}elle
sur laquelle les singularit\'{e}s $ADE$ sont localis\'{e}es.

\subsubsection{Spectre \`{a} quatre dimensions}

Avant d'aborder le spectre \`{a} quatre dimensions, commen\c{c}ons en
premier lieu par pr\'{e}ciser qu'\`{a} sept dimensions, le groupe de sym\'{e}%
trie global de la th\'{e}orie de super Yang-Mills provenant de la corde h%
\'{e}t\'{e}rotique sur $T^{3}$ est $SO(3)\times {SO(1,6)}.$ Le premier
facteur pr\'{e}sente la R-sym\'{e}trie tandis que le second n'est autre que
le groupe de Lorentz.\newline
Dans la repr\'{e}sentation adjointe du groupe de jauge, les champs de la th%
\'{e}orie se transforme comme suit: 
\begin{eqnarray}
{\text{Les champs de jauge}} &\longrightarrow &(1,7)  \nonumber \\
{\text{Les scalaires}} &\longrightarrow &(3,1)  \nonumber \\
{\text{Les fermions}} &\longrightarrow &(2,8).
\end{eqnarray}
De plus, les 16 supercharges se transforment aussi, 
\begin{equation}
16\longrightarrow (2,8).
\end{equation}
Pour $W$ arbitraire, le groupe de sym\'{e}trie dans ${W}\times {{\bf R}^{1,3}%
}$ se brise en 
\begin{equation}
SO(3)\times {SO(3)^{^{\prime }}\times {SO(1,3)}}.
\end{equation}
$SO(3)^{^{\prime }}$ est le groupe de structure du fibr\'{e} tangent \`{a} $%
W,$ alors que $SO(3)$ agit sur la fibre normale \`{a} $W$ dans $X.$ Dans ce
cas \`{a} quatre dimensions, les charges supersym\'{e}triques se
transforment en: 
\begin{equation}
(2,2,2)+(2,2,\overline{2}).
\end{equation}
Pour $D=4,$ la physique est d\'{e}crite par une th\'{e}orie de jauge qui
n'est pas supersym\'{e}trique puisque $W$ est une vari\'{e}t\'{e} courb\'{e}%
e de dimension 3. Mais localement, pour que ${{\bf C}^{2}}/{\Gamma _{ADE}} $
soit d'holonomie ${\bf G}_{2}$ il faut que cette th\'{e}orie soit supersym%
\'{e}trique. En effet ceci appara\^{\i}t clairement lorsque nous traitons la
structure-${\bf G}_{2}$. Pour cette raison, consid\'{e}rons tout d'abord la
structure $SU(2)$ de ${{\bf C}^{2}}/{\Gamma _{ADE}}.$ Cet espace \`{a}
quatre dimensions et d'holonomie $SU(2)$ est une vari\'{e}t\'{e} hyper-kahl%
\'{e}rienne qui admet trois 2-formes $\omega _{i}$ parall\`{e}les. Celles-ci
se transforment localement sous l'action de $SO(3).$ Ainsi dans ${{\bf C}^{2}%
},$ ces 2-formes sont explicitement donn\'{e}es par: 
\begin{eqnarray}
{\omega _{1}}+i{\omega _{2}} &=&du\wedge dv  \nonumber \\
{\omega _{3}} &=&\frac{i}{2}du\wedge {d{\overline{u}}}+dv\wedge {\ d{%
\overline{v}}}
\end{eqnarray}
et sont pr\'{e}serv\'{e}es par $\Gamma _{ADE}.$\newline
D'autre part, l'espace des modules des m\'{e}triques d'holonomie $SU(2)$ co%
\"{\i}ncide avec l'espace des modules de la th\'{e}orie de jauge poss\'{e}%
dant l'action $SO(3)$ (voir (\ref{e1})). Pour cette raison, le groupe de
rotation des 2-formes est identifi\'{e} avec le facteur $SO(3)$ de la th\'{e}%
orie de jauge \`{a} sept dimensions.\newline
Par ailleurs dans un rep\`{e}re plat ${e_{i}}$ de $W,$ la structure-${\bf G}%
_{2}$ de ${{\bf C}^{2}}/{\Gamma _{ADE}}$ prend la forme: 
\begin{equation}
{\varphi }=\frac{1}{6}{\omega _{i}}\wedge {e_{j}\delta ^{ij}}+{e_{1}\wedge {%
e_{2}}\wedge {e_{3}}}.
\end{equation}
Cette formule est invariante par $SO(3)$ \`{a} condition que ce groupe agit
sur les $e_{i}$ de la m\^{e}me fa\c{c}on qu'il agisse sur les $\omega _{i}.$
Cependant, $SO(3)^{^{\prime }}$ agit aussi sur les ${e_{i}}$ puisqu'il est
le groupe de structure du fibr\'{e} tangent de $W.$ Ainsi il doit \^{e}tre
identifi\'{e} avec $SO(3)$ afin que ${{\bf C}^{2}}/{\Gamma _{ADE}}$ admette
une m\'{e}trique d'holonomie ${\bf G}_{2}.$ \newline
Cette identification suscite la brisure des sym\'{e}tries en sous-groupe
diagonal $SO(3)^{\prime \prime }$ des deux $SO(3),$ incitant la th\'{e}orie 
\`{a} quatre dimensions \`{a} \^{e}tre supersym\'{e}trique. Apr\`{e}s cette
brisure, le groupe de sym\'{e}trie devient: 
\begin{equation}
{SO(3)^{\prime \prime }}\times {SO(1,3)}
\end{equation}
Dans ce cas, les champs se transforment comme suit: 
\begin{eqnarray}
{\text{Les champs de jauge}} &\longrightarrow &(3,1)+(1,4).  \nonumber \\
{\text{Les trois scalaires}} &\longrightarrow &(3,1).  \nonumber \\
{\text{Les supercharges}} &\longrightarrow &(1,2)+(3,2)+cc.  \nonumber \\
{\text{De m\^{e}me les fermions }} &\longrightarrow &(1,2)+(3,2)+cc.
\end{eqnarray}
Ainsi les champs scalaires, sous le groupe de Lorentz \`{a} $D=4$, forment
deux copies de la repr\'{e}sentation ${\bf 3}$ de $SO(3)^{\prime \prime }, $
autrement deux 1-formes de $W.$ A cet \'{e}gard, ils sont pr\'{e}cis\'{e}%
ment au nombre de $b_{1}(W)$. De plus leurs superpartenaires sont les $%
(3,2)+cc$ fermions. Ensemble, ces scalaires et ces fermions fournissent le
contenu de $b_{1}(W)$ supermultiplets chiraux. Cependant, le champ $(1,4)$
donne un champ de jauge \`{a} quatre dimensions dont les partenaires supersym%
\'{e}triques sont \'{e}videmment les fermions qui se transforment comme $%
(1,2)+cc$. Or puisque $X$ est d'holonomie ${\bf G}_{2},$ alors la vari\'{e}t%
\'{e} compacte $W$ doit v\'{e}rifier la condition $b_{1}(W)=0$. Par cons\'{e}%
quent, \`{a} quatre dimensions la physique de la th\'{e}orie-M au voisinage
des singularit\'{e}s $ADE$ est d\'{e}crite par la th\'{e}orie de super
Yang-Mills ${\cal N}=1$ pure.

\subsection{\ Mati\`{e}re chirale via la dualit\'{e} avec la corde h\'{e}t%
\'{e}rotique}

A ce stade nous avons montr\'{e} que les singularit\'{e}s orbifold $ADE$ de
la vari\'{e}t\'{e} d'holonomie ${\bf G}_{2}$ permettent d'avoir une th\'{e}%
orie avec un groupe de jauge non ab\'{e}lien. Mais ceci reste insuffisant
surtout que notre but majeur consiste \`{a} obtenir un mod\`{e}le de
physique des particules. Ce dernier exige la pr\'{e}sence de la mati\`{e}re
chirale charg\'{e}e sous les sym\'{e}tries de jauge dans le spectre \`{a}
quatre dimensions. Le type des singularit\'{e}s les plus simples fournissant
de la mati\`{e}re chirale sont les singularit\'{e}s coniques \cite%
{ref7,ref16}. \newline
Dans notre cas, la m\'{e}trique conique prend la forme: 
\begin{equation}
{{ds}^{2}}={{dr}^{2}}+{{{r}^{2}}g(Y)}
\end{equation}
o\`{u} $g(Y)$ n'est autre que la m\'{e}trique de la vari\'{e}t\'{e} $Y$ qui
est compacte de dimension six. Par suite la vari\'{e}t\'{e} $X$ poss\'{e}%
dant cette m\'{e}trique, est vue comme un c\^{o}ne sur $Y.$ De plus elle
dispose d'une singularit\'{e} \`{a} l'origine pour $r=0.$ Pour ce type de
singularit\'{e}, le spectre de la th\'{e}orie-M \`{a} $D=4$ contient bien de
la mati\`{e}re chirale.\newline

Il est clair \`{a} pr\'{e}sent que notre objectif consiste \`{a} construire
de fa\c{c}on explicite des vari\'{e}t\'{e}s avec singularit\'{e}s coniques.
Pour cette raison, nous allons une fois encore profiter des avantages de la
dualit\'{e} \`{a} sept dimensions avec la corde h\'{e}t\'{e}rotique sur $%
T^{3}$ que nous avons d\'{e}j\`{a} trait\'{e} pr\'{e}c\'{e}demment.

La corde h\'{e}t\'{e}rotique compactifi\'{e}e sur une vari\'{e}t\'{e} de
Calabi-Yau ${\cal Z}$ de dimension trois complexe, donne la mati\`{e}re
chirale dans son spectre \`{a} quatre dimensions \cite{ref17,ref18}. De
plus, selon un r\'{e}sultat de Strominger, Yau et Zaslow ${\cal Z}$ poss\'{e}%
de la propri\'{e}t\'{e} de la sym\'{e}trie miroir si elle est r\'{e}alis\'{e}%
e comme une fibration $T^{3}$ (avec singularit\'{e}s et monodromies) sur une
base $W$ (dans une limite de son espace des modules) \cite{ref19}. Ainsi, en
utilisant au niveau de chaque fibre $T^{3}$ la dualit\'{e} entre la corde h%
\'{e}t\'{e}rotique sur $T^{3}$ et la th\'{e}orie-M sur $K3,$ nous d\'{e}%
duisons gr\^{a}ce \`{a} l'argument adiabatique que la corde h\'{e}t\'{e}%
rotique sur ${\cal Z}$ est \'{e}quivalente \`{a} la th\'{e}orie-M sur une
vari\'{e}t\'{e} $X$ de dimension sept. Cette vari\'{e}t\'{e} X n'est autre
qu'une fibration $K3$ sur la m\^{e}me base $W$ (avec singularit\'{e}s et
monodromies ) et ayant une holonomie ${\bf G}_{2}.$\newline
Dans ce cas: 
\begin{eqnarray}
{\bf {\frac{\text{th\'{e}orie-M}}{{K3}\times{W}} \approx \frac{\text{h\'{e}t%
\'{e}rotique}}{{T^{3}}\times{W}}}}
\end{eqnarray}
Utilisons ce r\'{e}sultat afin de d\'{e}t\'{e}rminer dans $X$ le type de
singularit\'{e} suscitant de la mati\`{e}re chirale. Pour cette raison
supposons que la corde h\'{e}t\'{e}rotique sur ${\cal Z}$ poss\`{e}de une sym%
\'{e}trie de jauge non bris\'{e}e not\'{e}e $G$ et qui est simplement lac%
\'{e}e. Alors dans le contexte de $X,$ ceci signifie que chaque fibre $K3$
poss\`{e}de une singularit\'{e} de type $G$ et que, la famille des singularit%
\'{e}s dans $X$ est param\'{e}tris\'{e}e par la base $W.$\newline
Ainsi deux cas se distinguent:

\begin{enumerate}
\item Si l'espace normal \`{a} $W$, qui est r\'{e}guli\`{e}re, est une
famille des singularit\'{e}s de $G$ variant de fa\c{c}on r\'{e}guli\`{e}re,
alors dans ce cas la th\'{e}orie \`{a} basse \'{e}nergie est une th\'{e}orie
de jauge sur ${{\bf R}^{1,3}}\times {W}$ sans mati\`{e}re chirale.

\item En s'int\'{e}ressant au cas o\`{u} $W$ est plut\^{o}t singuli\`{e}re,
dans ce cas l'existence des multiplets chiraux est garantit par les
singularit\'{e}s de la base.
\end{enumerate}

Dans le but de d\'{e}terminer le type de ces singularit\'{e}s, nous
utilisons la dualit\'{e} avec la corde h\'{e}t\'{e}rotique. En particulier,
pla\c{c}ons-nous \`{a} titre d'exemple dans le cas de la corde h\'{e}t\'{e}%
rotique ${E_{8}}\times{E_{8}}$ et o\`{u} $G=SU(5)$ est sous-groupe de l'un
des $E_{8}$ \cite{ref7}. Ce mod\`{e}le contient de la mati\`{e}re chirale
dans la repr\'{e}sentation ${\it {\bf 5}}$ et ${\it {\bf 10}}$ du groupe $%
SU(5).$\newline
Etudions cet exemple sp\'{e}cifique de pr\'{e}s.

\subsubsection{{\protect\Large \ }Etude en language de la corde h\'{e}t\'{e}%
rotique}

Soit l'exemple de la repr\'{e}sentation ${\it {\bf 5}}.$ Le commutant de $%
SU(5)$ dans $E_{8}$ est une deuxi\`{e}me copie de $SU(5)$ not\'{e}e $%
SU(5)^{^{\prime }}.$ Or, puisque $SU(5)$ est non bris\'{e}, le groupe de
structure du fibr\'{e} de jauge $E$ de ${\cal Z}$ se r\'{e}duit de $E_{8}$ 
\`{a} $SU(5)^{^{\prime }}.$ De plus, la partie de la repr\'{e}sentation
adjointe de $E_{8}$ qui se transforme en ${\it {\bf 5}}$ sous $SU(5)$, se
transforme en ${\it {\bf 10}}$ sous $SU(5)^{^{\prime }}:$ 
\begin{eqnarray}
{E_{8}} &\longrightarrow &{{SU(5)}\times {SU(5)^{^{\prime }}}}  \nonumber \\
{{\text{partie de}}\ {\it {\bf 248}}} &\longrightarrow &{({\it {\bf 5}},{\it 
{\bf 10}})}.
\end{eqnarray}
D'autre part, le chiral non massif ${\it {\bf 5}}$ de $SU(5)$ s'obtient gr%
\^{a}ce \`{a} l'\'{e}quation de Dirac dans ${\cal Z}$ \`{a} valeurs dans $%
{\it {\bf 10}}$ de $SU(5)^{^{\prime }}.$ Effectivement, pour $SU(5)$ non bris%
\'{e}, ses repr\'{e}sentations non massives ${\it {\bf 5}}$ sont fournit par
les z\'{e}ros modes de l'\'{e}quation de Dirac.\newline
Par cons\'{e}quent, dans le cas des fibres $T^{3}$ de rayon $\xi $, avec $%
\xi $ tr\`{e}s petit, la r\'{e}solution de l'\'{e}quation de Dirac
s'effectue en deux \'{e}tapes:

\begin{enumerate}
\item Le long de la fibre.

\item Le long de la base.
\end{enumerate}

Autrement dit, l'op\'{e}rateur de Dirac ${\cal D}$ s'\'{e}crit: 
\begin{equation}
{\cal D}={{\cal D}_{T}}+{{\cal D}_{W}}
\end{equation}
o\`{u} ${\cal D}_{T}$ est l'op\'{e}rateur de Dirac le long de la fibre, par
contre ${\cal D}_{W}$ est l'op\'{e}rateur de Dirac le long de la base.%
\newline
La valeur propre de ${\cal D}_{T}$ entra\^{\i}ne un terme de ''masse''
effectif dans l'\'{e}quation de Dirac dans $W.$\newline
Dans le cas des fibres g\'{e}n\'{e}riques de ${\cal Z}\longrightarrow {W}$,
les z\'{e}ros modes sont localis\'{e}s au voisinage des points de $W$ sur
lesquels ${\cal D}_{T}$ admet un z\'{e}ro mode. En limitant l'\'{e}tude \`{a}
une seule fibre $T^{3},$ le $SU(5)^{^{\prime }}$ de $E$ est d\'{e}crit comme
fibr\'{e} plat avec des monodromies autour des trois directions du tore. Par
ailleurs au-dessus de quelques points $P$ de $W,$ un z\'{e}ro mode de ${\cal %
D}_{T}$ surgit pr\'{e}cis\'{e}ment si les monodromies dans la fibre sont
tous triviales pour certains vecteurs de la repr\'{e}sentation ${\it {\bf 10}%
}.$ Ceci signifie que les monodromies du sous-groupe $SU(5)^{^{\prime }}$
laissent ces vecteurs fixes.\newline
Si ${\it {\bf 10}}$ est repr\'{e}sent\'{e}e par une matrice antisym\'{e}%
trique $(${\it {\bf 5}}$,${\it {\bf 5}}$)$ que nous notons $B^{ij}$ avec $%
i,j=1,\ldots ,5,$ alors le vecteur invariant par la monodromie correspond 
\`{a} une matrice $B$ ayant quelques \'{e}l\'{e}ments non nuls. Il est clair 
\`{a} pr\'{e}sent que le sous-groupe de $SU(5)^{^{\prime }}$ sous lequel $B$
est invariant, est un sous-groupe de ${SU(2)}\times {SU(3)}$ (tel que $SU(2)$
agit sur les coordonn\'{e}es dont $i,j=1,2,$ cependant $SU(3)$ agit sur $%
i,j=3,4,5$).\newline
Particuli\`{e}rement, nous choisissons une base qui diagonalise les
monodromies au voisinage de $P$ et qui d'autre part, entrave $B^{12}$ \`{a} 
\^{e}tre le seul \'{e}l\'{e}ment non nul de la matrice $B.$ Dans ce cas, le
sous-groupe de $SU(5)^{^{\prime }}$ laissant $B$ fixe est exactement ${SU(2)}%
\times {SU(3)}.$ Or, le commutant de ce dernier dans $E_{8}$ n'est autre que 
$SU(6).$ Ainsi, nous distinguons deux cas:\newline

\underline{Au dessus du point{\bf \ }$P$}{:}

\begin{enumerate}
\item Les monodromies commutent non seulement avec $SU(5)$ mais aussi avec $%
SU(6).$

\item Les monodromies attribuent de grandes masses \`{a} tout les modes de $%
E_{8}$ except\'{e} ceux de la repr\'{e}sentation adjointe de $SU(6).$

\item Le groupe de jauge se restreint \`{a} $SU(6)$ au lieu de $E_{8}.$
\end{enumerate}

\underline{Loin du point $P$}{\bf :}

\begin{enumerate}
\item Les monodromies brisent $SU(6)$ en ${SU(5)}\times {U(1)}.$
\end{enumerate}

Par cons\'{e}quent, r\'{e}duire l'\'{e}tude de $E_{8}$ \`{a} $SU(6)$ revient 
\`{a} consid\'{e}rer $U(1)$ comme fibr\'{e} de jauge plut\^{o}t que $%
SU(5)^{^{\prime }}.$\newline
Autrement, 
\begin{eqnarray}
{E_{8}} &\longrightarrow &{SU(6)}  \nonumber \\
{\text{implique que:}}\ {SU(5)^{^{\prime }}} &\longrightarrow &{U(1)},
\end{eqnarray}
surtout que $U(1)$ est le second facteur dans ${{SU(5)}\times {U(1)}}\subset
SU(6).$\newline
Or dire que $SU(6)$ est non bris\'{e} au dessus de $P$ pour la corde h\'{e}t%
\'{e}rotique, signifie dans le contexte de la th\'{e}orie-M que la fibre sur 
$P$ poss\`{e}de une singularit\'{e} $SU(6).$ Par contre la non-brisure de ${%
SU(5)}\times {U(1)}$ loin de $P,$ s'explique au niveau de la th\'{e}orie-M
par le fait que la fibre g\'{e}n\'{e}rique contient seulement la singularit%
\'{e} $SU(5)$ au lieu de $SU(6).$ Et de plus le $U(1)$ non bris\'{e} est port%
\'{e} par la 3-forme $C$ de la th\'{e}orie-M.\newline
En conclusion, la mati\`{e}re chirale provenant de la corde h\'{e}t\'{e}%
rotique sur ${\cal Z}$ est localis\'{e}e aux points $P$ de $W$ o\`{u} les
monodromies dans les fibres $T^{3}$ sont triviales. \newline
Nous allons compl\'{e}ter cette \'{e}tude par passage \`{a} la th\'{e}orie-M
dans ce qui suit.

\subsubsection{{\protect\Large \ }Etude en language de la th\'{e}orie-M}

Avant de reprendre la description faite dans le cadre de la th\'{e}orie-M
sur une vari\'{e}t\'{e} $X$ d'holonomie ${\bf G}_{2},$ commen\c{c}ons tout
d'abord par traiter bri\`{e}vement le cas de l'ing\'{e}nierie g\'{e}om\'{e}%
trique de la mati\`{e}re charg\'{e}e pour type IIA sur une Calabi-Yau de
dimension trois complexe \cite{ref20}. Cette vari\'{e}t\'{e} not\'{e}e $%
{\cal R}$ est r\'{e}alis\'{e}e comme une fibre $K3$ sur une base $%
Q^{^{\prime }}$ de sorte que:

\begin{enumerate}
\item Sur un point distingu\'{e} $P\in Q^{^{\prime }},$ la singularit\'{e}
est de type $\widehat{G}.$

\item Sur un point g\'{e}n\'{e}rique de $Q^{^{\prime }},$ cette singularit%
\'{e} est remplac\'{e}e par une de type $G$ o\`{u}: 
\begin{equation}
{\text{r}ang}{\widehat{G}}=1+{\text{r}ang}G
\end{equation}
\end{enumerate}

A ce niveau, nous restreignons notre \'{e}tude \`{a} ${\widehat{G}}=SU(6)$
et $G=SU(5)$ puisque c'est le cas que nous avons d\'{e}j\`{a} trait\'{e}
dans le cadre de la corde h\'{e}t\'{e}rotique.\newline
La singularit\'{e} $SU(6)$ est d\'{e}crite par l'\'{e}quation: 
\begin{equation}
xy=z^{6}.
\end{equation}
Or, son d\'{e}pliement d\'{e}pend de cinq param\`{e}tres complexes et s'\'{e}%
crit: 
\begin{equation}
xy=z^{6}+P_{4}(z)
\end{equation}
o\`{u} $P(z)$ est un polyn\^{o}me quartique en $z$.\newline
Dans le but de d\'{e}former la singularit\'{e} $SU(6)$ en retenant la
singularit\'{e} $SU(5),$ le polyn\^{o}me $z^{6}+P_{4}(z)$ doit avoir une
racine du $5^{\text{\`{e}}me}$ ordre. Sous cette condition, la d\'{e}%
formation prend la forme: 
\begin{equation}
xy=(z+5\varepsilon ){(z-\varepsilon )^{5}}  \label{b1}
\end{equation}
o\`{u} $\varepsilon $ est un param\`{e}tre complexe de la base $Q^{^{\prime
}}.$\newline
En plus du fait qu'elle d\'{e}crit le d\'{e}pliement partiel de la singularit%
\'{e} $SU(6),$ cette \'{e}quation (\ref{b1}) donne aussi la structure
complexe de l'espace total ${\cal R}.$ \newline
Dans la suite, nous allons prolonger cette \'{e}tude \`{a} la vari\'{e}t\'{e}
$X$ d'holonomie ${\bf G}_{2}$ qui nous int\'{e}resse \cite{ref17,ref18}.
Dans ce cas, la construction est similaire \`{a} la seule diff\'{e}rence,
est d'aborder la singularit\'{e} $SU(6)$ en tant qu'une vari\'{e}t\'{e}
hyperkahl\'{e}rienne au lieu de complexe comme est le cas pour cet exemple
de ${\cal R}.$ Pour cette raison, chaque param\`{e}tre complexe de $P_{4}$
doit \^{e}tre accompagn\'{e} par un param\`{e}tre r\'{e}el qui contr\^{o}le
l'aire d'un diviseur exceptionnel lors de la d\'{e}formation de la singularit%
\'{e}. Par cons\'{e}quent, les param\`{e}tres ne sont plus cinq complexes
mais plut\^{o}t une famille de cinq triplets de param\`{e}tres r\'{e}els.
L'espace des param\`{e}tres de cette d\'{e}formation n'est autre que $W$ et
l'espace total est la vari\'{e}t\'{e} singuli\`{e}re d'holonomie ${\bf G}%
_{2} $ dont la singularit\'{e} produit la mati\`{e}re chirale.\newline
A ce niveau, il est \'{e}vident de se demander comment sera le d\'{e}%
pliement dans ce cas?\newline

Afin de mieux illustrer cette id\'{e}e, nous ferons recours \`{a} la
description de Kronheimer du d\'{e}pliement via le quotient hyperkahl\'{e}%
rien. De plus, au lieu de limiter l'\'{e}tude comme nous l'avons fait jusqu'%
\`{a} pr\'{e}sent \`{a} la singularit\'{e} $SU(6),$ nous allons plut\^{o}t g%
\'{e}n\'{e}raliser l'analyse en consid\'{e}rant le d\'{e}pliement de $%
SU(N+1) $ qui garde la singularit\'{e} $SU(N).$ \newline
Ce nouveau d\'{e}pliement est appel\'{e} d\'{e}pliement hyperkahl\'{e}rien,
de plus il s'obtient en prenant un syst\`{e}me de $N+1$ hypermultiplets $%
\phi _{0},$$\phi _{1},$ $\cdots ,$$\phi _{N}$ avec une action de $K={U(1)^{N}%
}.$ Sous le $i^{\text{\`{e}}me}$ $U(1)$ (o\`{u} $i=1,\ldots ,N$), $\phi _{i}$
a la charge $1$ et $\phi _{i-1}$ porte la charge $-1$ par contre tout les
autres sont neutres. Ainsi ces hypermultiplets param\'{e}trisent le produit
de $N+1$ copies de ${\bf R}^{4}$ que nous notons ${\bf H}^{N+1}$ pour ${\bf H%
}\simeq {{\bf R}^{4}}.$ \newline
Par ailleurs, le quotient hyperkahl\'{e}rien de ce ${\bf H}^{N+1}$ par $K$
se d\'{e}finit en deux \'{e}tapes:

\begin{enumerate}
\item En posant les $\overrightarrow{D}$-champs \'{e}gaux \`{a} z\'{e}ro.

\item En divisant par $K.$
\end{enumerate}

Ce qoutient est alors not\'{e} par ${{\bf H}^{N+1}}//{K}$ et il est
isomorphe \`{a} ${\bf H}/{{\bf Z}_{N+1}}$ de singularit\'{e} $SU(N+1).$%
\newline
En global, le d\'{e}pliement contient $3N$ param\`{e}tres puisque $%
\overrightarrow{D} $ poss\`{e}de pour chaque $U(1)$ trois composantes dont
le groupe de rotation est $SU(2)$ (R-sym\'{e}trie). D'autre part, le d\'{e}%
pliement partiel de $SU(N+1),$ tout en retenons la singularit\'{e} $SU(N),$
s'obtient alors:

\begin{enumerate}
\item En imposant aux $3(N-1)$ param\`{e}tres \`{a} \^{e}tre \'{e}gaux \`{a}
z\'{e}ro.

\item En laissant les trois param\`{e}tres qui restent varier. Notons que
ces trois ne sont autre que les valeurs de $\overrightarrow{D}$ pour l'un
des $U(1).$
\end{enumerate}

Pour effectuer cette proc\'{e}dure, commen\c{c}ons d'abord par r\'{e}aliser $%
K$ comme 
\begin{equation}
K={K^{^{\prime}}}\times {U(1)^{^{\prime}}}
\end{equation}
o\`{u} $U(1)^{^{\prime}}$ est l'un des facteurs choisis de $K= U(1)^{N}.$
Ensuite, prenons le quotient hyperkahl\'{e}rien de ${\bf H}^{N+1}$ par $%
K^{^{\prime}}$ pour avoir une vari\'{e}t\'{e} hyperkahl\'{e}rienne de
dimension $8$ not\'{e}e: 
\begin{equation}
\widehat{X}={{\bf H}^{N+1}}//{K^{^{\prime}}}.
\end{equation}
Apr\`{e}s, consid\'{e}rons le quotient ordinaire et non pas le quotient
hyperkahl\'{e}rien de $\widehat{X}$ par $U(1)^{^{\prime}}.$ Ce quotient va
nous fournir la vari\'{e}t\'{e} $X$ de dimension sept qui admet pour groupe
d'holonomie, le groupe ${\bf G}_{2}$: 
\begin{equation}
X={\widehat{X}}/{U(1)^{^{\prime}}}.
\end{equation}

De plus, afin d'identifier les groupes d'action sur ${\bf H}$ et ${\bf H}%
^{^{\prime}},$ param\'{e}trisons ${\bf H}$ et ${\bf H}^{^{\prime}}$
respectivement par les paires des variables complexes $(a,b)$ et $%
(a^{\prime},b^{\prime}).$ \newline
Alors:

\begin{enumerate}
\item L'action de ${\bf Z}_{N}$ sur ${\bf H}:$%
\begin{equation}
\left( 
\begin{array}{c}
a \\ 
b%
\end{array}
\right) \qquad \rightarrow \qquad \left( 
\begin{array}{c}
{e^{\frac{2ik\pi }{N}}}a \\ 
{e^{\frac{-2ik\pi }{N}}}b%
\end{array}
\right)
\end{equation}

\item L'action de $U(1)^{^{\prime }},$ commutant avec ${\bf Z}_{N}$ et pr%
\'{e}servant la structure hyperkahl\'{e}rienne, est:\newline
$\bullet $ {sur ${\bf H}$} 
\begin{equation}
\left( 
\begin{array}{c}
a \\ 
b%
\end{array}
\right) \qquad \rightarrow \qquad \left( 
\begin{array}{c}
{e^{\frac{i\psi }{N}}}a \\ 
{e^{\frac{-i\psi }{N}}}b%
\end{array}
\right)
\end{equation}
$\bullet $ {sur ${\bf H}^{^{\prime }}$} 
\begin{equation}
\left( 
\begin{array}{c}
a^{\prime } \\ 
b^{\prime }%
\end{array}
\right) \qquad \rightarrow \qquad \left( 
\begin{array}{c}
{e^{-i\psi }}a^{\prime } \\ 
{e^{i\psi }}b^{\prime }%
\end{array}
\right)
\end{equation}
\end{enumerate}

Enfin, en posant 
\begin{eqnarray}
\lambda&=&{e^{i\psi}}  \nonumber \\
\sigma_{1}= {\overline{a}^{\prime}}& {\text et}& \sigma_{2}= {b^{\prime}} 
\nonumber \\
\sigma_{3}= {a} & {\text et}& \sigma_{4}= {\overline {b}},
\end{eqnarray}
le quotient $\left({{\bf H}/{{\bf Z}_{N}}}\times {{\bf H}^{^{\prime}}}%
\right)/{U(1)^{^{\prime}}}$ peut \^{e}tre d\'{e}crit gr\^{a}ce \`{a} ces
quatre variables complexes modulo l'\'{e}quivalence: 
\begin{eqnarray}
\left(\sigma_{1},\sigma_{2},\sigma_{3},\sigma_{4}\right)&\longrightarrow&
\left( {\lambda^{N}}\sigma_{1},{\lambda^{N}}\sigma_{2},{\lambda}\sigma_{3},{%
\lambda}\sigma_{4}\right)  \nonumber \\
|\lambda|&=&1.
\end{eqnarray}
Ce quotient nous d\'{e}crit alors un c\^{o}ne sur l'espace projectif pond%
\'{e}r\'{e} ${\bf WCP}^{3}_{N,N,1,1}.$\newline
Par cons\'{e}quent, nous d\'{e}duisons que la vari\'{e}t\'{e} d'holonomie $%
{\bf G}_{2}$ que nous cherchons est r\'{e}alis\'{e}e en tant qu'un c\^{o}ne
sur l'espace projectif pond\'{e}r\'{e}. De plus, \'{e}tant donn\'{e} que $%
{\bf WCP}^{3}_{N,N,1,1}$ dispose d'une famille de singularit\'{e}s $A_{N-1}$
aux points $(\omega_{1},\omega_{2},0,0),$ qui appara\^{\i}t clairement en
posant $\lambda= e^{\frac{2i\pi}{N}},$ alors notre vari\'{e}t\'{e} poss\'{e}%
de bien \`{a} son tour une famille de singularit\'{e}s $A_{N-1}.$

\section{Conclusion}

Loin de chercher l'exhaustivit\'{e}, ce travail a tent\'{e} d'introduire la
th\'{e}orie-M qui est envisag\'{e}e susceptible d'unifier les cinq mod\`{e}%
les des th\'{e}ories des supercordes, diff\'{e}rentes dans leurs
descriptions perturbatives faiblement coupl\'{e}es.\newline
Nous avons \'{e}galement essay\'{e} de montrer dans ce papier, que la
compactification de la th\'{e}orie-M, supergravit\'{e} \`{a} onze dimensions
avec 32 supercharges, sur une vari\'{e}t\'{e} de dimension sept et
d'holonomie ${\bf G}_{2}$ fournit une th\'{e}orie \`{a} quatre dimensions de
supersym\'{e}trie ${\cal N}=1.$\newline
Nous avons enfin proc\'{e}d\'{e} \`{a} l'\'{e}tude de deux cas:

\begin{enumerate}
\item D'une part, lorsque $X$ \'{e}tait r\'{e}guli\`{e}re, l'analyse de
Kaluza-Klein des champs \`{a} onze dimensions nous a produit \`{a} quatre
dimensions, une th\'{e}orie de supergravit\'{e} ${\cal N}=1$ coupl\'{e}e 
\`{a} $b_{2}(X)$ multiplets vectoriels ab\'{e}liens et $b_{3}(X)$ multiplets
chiraux.

\item D'autre part, dans le cas o\`{u} $X$ poss\'{e}dait un certain type de
singularit\'{e}s, la dynamique \`{a} basse \'{e}nergie a \'{e}t\'{e} d\'{e}%
crite par une th\'{e}orie de jauge non ab\'{e}lienne coupl\'{e}e \`{a} de la
mati\`{e}re chirale. En particulier, les sym\'{e}tries de jauge non ab\'{e}%
liennes locales ont \'{e}t\'{e} obtenues quand $X$ a \'{e}t\'{e} r\'{e}alis%
\'{e}e comme une $K3$ fibration sur une base de dimension trois. Tandis que
des singularit\'{e}s additionnelles, plus pr\'{e}cis\'{e}ment, les singularit%
\'{e}s coniques de codimension sept dans $X$ nous ont fournit de la mati\`{e}%
re chirale.
\end{enumerate}

{\bf {Acknowledgement}}: Je tiens \`{a} remercier Prof. E. H. Saidi ainsi
que Dr. A. Belhaj pour les discussions. Je remercie \'{e}galement L'ICTP de
Trieste de m'avoir faite profiter de ses s\'{e}minaires..

\end{document}